\documentclass[a4paper,11pt]{article}
\usepackage{jcappub}
\usepackage{lineno}

\usepackage{graphicx}
\graphicspath{{Figures/}}
\usepackage{float}
\usepackage{physics}
\usepackage{wrapfig}
\usepackage[utf8x]{inputenc}
\usepackage{amsmath}
\usepackage{amssymb}
\usepackage{orcidlink}
\usepackage{amsthm}
\usepackage{mathtools}
\usepackage{color}
\usepackage{multirow}
\usepackage{tabularx}
\usepackage{comment}
\usepackage{soul}
\usepackage{siunitx}
\usepackage{booktabs}

\DeclareSIUnit{\bar}{bar}
\DeclareSIUnit{\dBm}{dBm}
\DeclareSIUnit{\sample}{Sa}

\usetikzlibrary{arrows.meta, positioning}

\begin{document}
\makeatletter
\gdef\@fpheader{}
\makeatother

\title{ADAMOS: Axion Daily Modulation Searches for Dark Matter at \SI{20}{\GHz}}

\author{Marios Maroudas\orcidlink{0000-0003-1294-1433}}
\affiliation{Institut für Experimentalphysik, Universität Hamburg, Luruper Chaussee 149, D-22761 Hamburg, Germany}
\emailAdd{marios.maroudas@uni-hamburg.de}

\author{Toma-Stefan Cezar\orcidlink{0009-0009-5121-1821}}
\affiliation{Institut für Experimentalphysik, Universität Hamburg, Luruper Chaussee 149, D-22761 Hamburg, Germany}

\author{Antonios Gardikiotis\orcidlink{0000-0002-4435-2695}}
\affiliation{National Centre of Scientific Research "Demokritos", Patr. Gregoriou E \& 27 Neapoleos Str, 15341 Agia Paraskevi, Greece}

\author{Dieter Horns\orcidlink{0000-0003-1945-0119}}
\affiliation{Institut für Experimentalphysik, Universität Hamburg, Luruper Chaussee 149, D-22761 Hamburg, Germany}

\date{\today}

\abstract{
The ADAMOS (Axion Daily Modulation Searches) project aims to explore the nature of dark matter (DM) through a novel axion haloscope experiment. We propose a fixed-frequency cavity resonator operating at \SI{20}{\GHz} at the University of Hamburg, using an innovative “thin-shell” design that preserves a large detection volume at high frequencies. The experiment will be installed in an existing \SI{14}{\tesla} superconducting magnet and connected to a highly sensitive RF chain with continuous in situ calibration to eliminate temperature-dependent gain drifts, constituting an essential improvement based on lessons learned from previous attempts. ADAMOS will conduct simultaneous searches for three classes of axion signals: (1) conventional cold DM axions, (2) relativistic axions from axion quark nugget annihilations exhibiting daily modulations, and (3) transient enhancements from streaming DM. By targeting this unexplored frequency regime with a robust, calibrated, and versatile setup, ADAMOS will open new discovery channels in a previously unexplored region of the dark sector.
}


\maketitle
\flushbottom

\section{Introduction}
\label{section:Introduction}

The quest to understand the nature of Dark Matter (DM) remains one of the most compelling challenges in modern physics. Despite overwhelming astrophysical and cosmological evidence for its existence, DM has remained elusive in direct experimental searches. Among the leading candidates, axions provide a natural solution to the strong CP problem in quantum chromodynamics (QCD) \cite{Peccei_1977_constraints} and can also act as Cold Dark Matter (CDM) \cite{Sikivie_1983_experimental}. One of the most established experimental techniques to detect axions is the haloscope approach pioneered by Sikivie \cite{Sikivie_1983_experimental}, which employs a high-quality-factor microwave cavity immersed in a strong magnetic field to resonantly convert axions into detectable photons.

Traditional haloscope searches such as ADMX \cite{ADMX_2021_search}, CAPP \cite{CAPP_2024_ahn}, and CAST-CAPP \cite{Adair_CAST-CAPP_2022} are optimized for the narrow-band ($\Delta\nu/\nu \sim 10^{-6}$) signal expected from a virialized halo. At higher frequencies (larger axion masses), the cavity volume decreases rapidly as the resonant wavelength becomes smaller, making high-frequency sensitivity extremely challenging and leaving the region above $\sim$\,\SI{10}{\GHz} largely unexplored. This region is nevertheless well motivated by recent lattice QCD calculations for the QCD axion mass \cite{Borsanyi_2016_mass}, highlighting a critical gap in the experimental landscape.

Beyond the standard CDM paradigm, alternative axion production mechanisms predict distinct and non-standard signatures that narrow-band searches may miss. The Axion Quark Nugget (AQN) model \cite{Zhitnitsky_2003_aqn} proposes that DM consists of macroscopic nuggets of nuclear matter stabilized by an axion domain wall (ADW). In this framework, mildly relativistic axions with a large bandwidth ($\Delta\nu/\nu \sim 1$) are emitted during annihilations as AQNs traverse the Earth. Additionally, their flux is modulated with a sidereal day period due to Earth’s rotation relative to the galactic DM wind \cite{Liang_aqn_2020, Budker_AQN_2020}. A previous search by the CAST-CAPP collaboration \cite{Caspers_daily_2025} demonstrated the feasibility of this detection approach, but was ultimately limited by temperature-dependent gain drifts in the readout chain. More recently, the CAPP collaboration \cite{CAPP_2025_first_aqn} performed a dedicated AQN daily modulation search, establishing new exclusion limits in the \SI{1.9}{\micro\eV} to \SI{9.3}{\micro\eV} mass range. Axion-related daily modulation signatures have also been searched for in other frameworks \cite{ADMX_2023_background} with similar stability limitations.

Furthermore, a significant fraction of the halo may reside in substructure such as streams or miniclusters \cite{Vogelsberger_2011_streams, Tkachev_1991_minicluster}. Slow-moving axions can then be transiently enhanced by gravitational focusing from solar system bodies \cite{Kryemadhi_gravitational_2023, OHare_2023_Minicluster}, leading to temporally localized and strongly boosted local axion densities. These short-duration, narrowband excesses are naturally averaged out in conventional long-integration analyses. A dedicated search for such transient events requires a high-duty-cycle experiment with time-domain analysis capabilities and high-resolution FFTs.

In this paper, we present the design and projected sensitivity of ADAMOS (Axion DAily MOdulation Searches), a novel fixed-frequency haloscope operating around \SI{20}{\GHz}. ADAMOS introduces three key innovations: (1) a “thin-shell’’ cavity geometry \cite{Kuo_2019_large} that preserves a large ($\sim$\SI{1}{\litre}) detection volume at high frequency; (2) a continuously calibrated, highly stable RF chain to suppress temperature-driven gain drifts; and (3) a unified DAQ and analysis framework capable of performing three simultaneous searches for CDM axions, daily-modulated relativistic axions from AQNs, and transient signals from streaming DM. By targeting this unexplored high-frequency regime with a robust, calibrated, multi-channel approach, ADAMOS will open new discovery windows in the dark sector and establish the first state-of-the-art cavity haloscope of its kind in Germany.

\section{Theoretical Motivation and Signal Models}
\label{sec:theory}

The ADAMOS experiment is designed to probe three distinct, yet complementary, signatures of axion DM. These correspond to different production mechanisms and astrophysical distributions, each yielding a unique experimental fingerprint in a microwave cavity detector. This section details the theoretical underpinnings of the conventional CDM axion signal, the predicted daily modulation from AQNs, and the potential transient signals from DM streams.

\subsection{Cold Dark Matter Axion Signal}
\label{subsec:cdm_axions}

\begin{figure}[htb!]
    \centering
    \includegraphics[width=0.7\columnwidth]{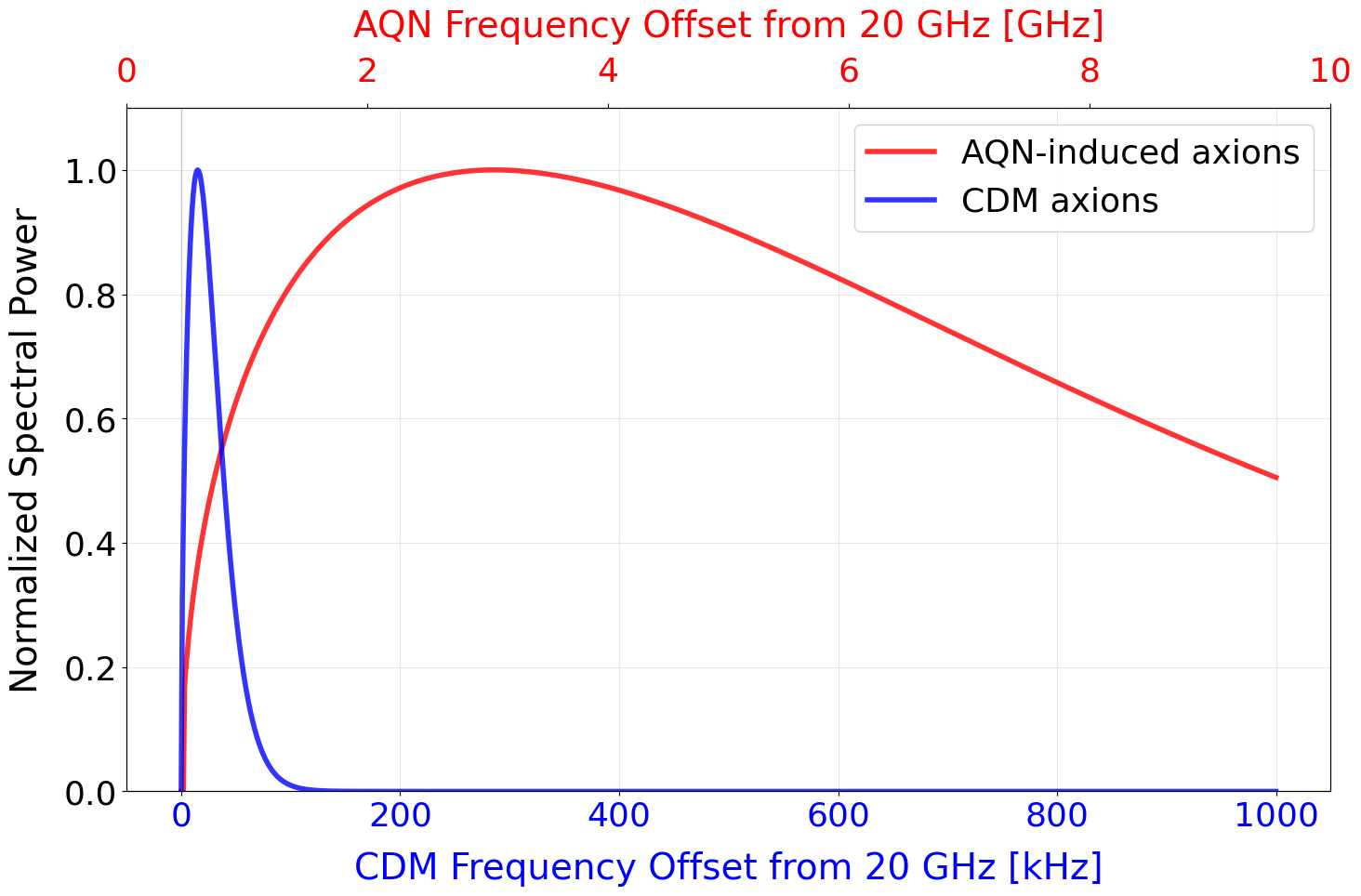}
    \caption{Comparison of AQN-induced (\textcolor{red}{red}) and CDM (\textcolor{blue}{blue}) axion spectral shapes at \SI{20}{\GHz}. The AQN signal is relativistic and broadband ($\Delta f/f \sim 1$), while the CDM signal is a boosted Maxwellian with a FWHM of $\sim\SI{36}{\kHz}$ ($\Delta f/f \sim 10^{-6}$), accounting for terrestrial motion. Note the dual frequency scales: the bottom axis covers \SI{1}{\MHz} for CDM, while the top axis spans \SI{10}{\GHz} for AQN.}
    \label{fig:cdm_aqn_signal_shape}
\end{figure}

The QCD axion provides an elegant solution to the strong CP problem \cite{Peccei_1977_constraints} and is a well-motivated CDM candidate \cite{Sikivie_1983_experimental}. In the presence of a strong magnetic field, these galactic halo axions can convert into photons via the inverse Primakoff effect, with a power given by:
\begin{equation}
P_{a \rightarrow \gamma} = g_{a\gamma\gamma}^2 \frac{\rho_a}{m_a} B_0^2 V C Q_L.
\end{equation}
where $ g_{a\gamma\gamma}$  is the axion photon coupling, $\rho_a$ is the local DM density, which is assumed to be comprised of axions, $m_a$ is the axion mass, $B_0$
is the magnetic field strength, $V$ is the cavity volume, 
$C$ is the cavity form factor, and $Q_L$ is the loaded quality factor of the cavity. The signal is quasi-monochromatic, with a fractional bandwidth of $\Delta\nu/\nu \sim 10^{-6}$, set by the virial velocity distribution of the galactic halo. The expected spectral line in the lab frame follows the Standard Halo Model, i.e., a Maxwell–Boltzmann velocity distribution boosted by the laboratory velocity \cite{turner_1990_periodic, haystac_2017_analysis}. The blue line in Fig.~\ref{fig:cdm_aqn_signal_shape} illustrates the resulting spectral shape of CDM axions around \SI{20}{\GHz} for a detector located at the University of Hamburg (latitude \SI{53.5}{\degree}) with a characteristic linewidth of approximately \SI{36}{\kHz}. This width accounts for the laboratory velocity boost, which broadens the signal beyond the naive $10^{-6}$ fractional width.

Recent advances in lattice QCD simulations suggest that the QCD axion mass would plausibly lie in the \SIrange{10}{100}{\micro\eV} range (i.e.\ frequencies in the \SIrange{2}{20}{\GHz}) \cite{Borsanyi_2016_mass}. However, at higher frequencies, the cavity volume decreases rapidly as the resonant wavelength becomes smaller, which causes a strong loss in axion-to-photon conversion sensitivity. For instance, a cylindrical cavity designed for the fundamental TM$_{010}$ mode at \SI{20}{\GHz} would have a radius of only $\sim$\SI{5.7}{\milli\meter}, directly impacting the achievable signal power.

To compensate for this sensitivity limitation at high frequencies, recent efforts have either focused on increasing the signal power (e.g.\ by increasing the quality factor $Q_L$ via superconducting coatings \cite{capp_2022_hts}) or on developing new approaches which can reach higher axion masses without suffering the same scaling limitations. Examples include multi-cell cavity concepts \cite{Goryachev_2017_arrays, rades_2020_scalable, Junu_2023_multicell}, dielectric-based haloscopes \cite{horns_2013_dish, madmax_2017_dielectric}, plasma-based haloscopes \cite{millar_2023_alpha}, and alternative cavity geometries \cite{McAllister_2017_organ, Kuo_2019_large, Withers_2024_beehive}. These developments highlight the strong experimental need for resonator concepts capable of maintaining large effective detection volumes while accessing high axion masses.

\subsection{Daily Modulation from Axion Quark Nuggets}
\label{subsec:aqn_modulation}

\subsubsection{Axion Quark Nugget Model}
\label{subsubsec:AQN_model}

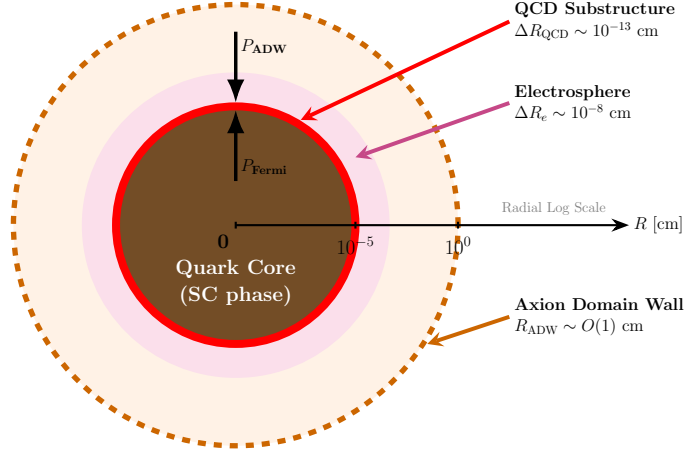
\begin{figure}[!htb]
    \centering
    \begin{tikzpicture}[font=\sffamily\large, scale=0.45, transform shape]

    %
    %

    \fill[orange!10] (0,0) circle (6.5cm);
    \draw[orange!80!black, dashed, line width=2pt] (0,0) circle (6.5cm);
    
    \fill[magenta!15] (0,0) circle (4.5cm);
    
    \fill[brown!60!black] (0,0) circle (3.5cm);
    \draw[red, line width=3pt] (0,0) circle (3.5cm); 

    \draw[Latex-, line width=1.3pt] (0, 3.6) -- (0, 5.7) 
        node[pos=0.8, right, font=\Large\bfseries] {$P_{\text{ADW}}$};
        
    \draw[-Latex, line width=1.3pt] (0, 1.3) -- (0, 3.4) 
        node[pos=0.2, right, font=\Large\bfseries] {$P_{\text{Fermi}}$};

    \node[white, align=center, font=\LARGE\bfseries] at (0, -1.7) {Quark Core\\(SC phase)};

    \draw[thick, -Stealth] (0,0) -- (11.5,0) node[anchor=west, font=\Large] {$R$ [cm]};
    
    \draw[thick] (0, 0.1) -- (0, -0.1) node[below left=2pt, font=\LARGE\bfseries] {0};
    \draw[thick] (3.5, 0.1) -- (3.5, -0.1) node[below=3pt, font=\LARGE\bfseries] {$10^{-5}$};
    \draw[thick] (6.5, 0.1) -- (6.5, -0.1) node[below=3pt, font=\LARGE\bfseries] {$10^{0}$};
    
    \node[font=\large, gray] at (9.3, 0.5) {Radial Log Scale};

    \draw[stealth-, red, line width=1.5pt] (1.9, 3.1) -- (8.0, 6) 
        node[right, black, align=left, font=\Large] {\textbf{QCD Substructure}\\ $\Delta R_{\text{QCD}} \sim 10^{-13}$ cm};

    \draw[stealth-, magenta!80!black, line width=1.5pt] (3.5, 2) -- (8.0, 3.6) 
        node[right, black, align=left, font=\Large] {\textbf{Electrosphere}\\ $\Delta R_e \sim 10^{-8}$ cm};

    \draw[stealth-, orange!80!black, line width=1.5pt] (5.6, -3.5) -- (8.0, -2.7) 
        node[right, black, align=left, font=\Large] {\textbf{Axion Domain Wall}\\ $R_{\text{ADW}} \sim O(1)$ cm};

    \end{tikzpicture}
    \caption{Schematic internal structure of an AQN. The color superconducting (SC) quark core ($R \sim 10^{-5}$ cm, $B \sim 10^{25}$), in brown, is bounded by a thin QCD substructure, in red, ($\Delta R_{\text{QCD}} \sim 10^{-13}$ cm) and an electrosphere ($\Delta R_e \sim 10^{-8}$ cm), in pink. The nugget is stabilized by the pressure of an ADW in gold, where hydrostatic equilibrium is maintained by the balance of $P_{\text{ADW}}$ and $P_{\text{Fermi}}$. For an axion mass of $m_a \approx \SI{82.7}{\micro\eV}$ the characteristic radius $R_{\text{ADW}} \propto m_a^{-1} \sim O(1)~\si{\cm}$. The radial reference axis is logarithmic in the absolute radius $R$ to illustrate the large hierarchy between the microscopic core and the macroscopic ADW. The QCD substructure and electrosphere are shown as finite layers for visual clarity, while their thicknesses $\Delta R \ll R$ are not drawn to scale. Adapted from \cite{Zhitnitsky_2023_structure}.}
    \label{fig:aqn_structure} 
\end{figure}

The Axion Quark Nugget (AQN) model offers an alternative DM paradigm where the similarity between the DM and visible matter densities, $\Omega_\text{dark} \sim \Omega_\text{visible}$, emerges naturally \cite{Zhitnitsky_2003_aqn}. In this framework, DM consists of macroscopic nuggets of baryonic matter in a color superconducting (CS) phase, confined by an ADW. The multi-layered internal structure of such a nugget, including the various physical scales involved, is illustrated in Fig.~\ref{fig:aqn_structure}. For a representative axion mass $m_a \approx \SI{82.7}{\micro\eV}$, corresponding to a frequency of approximately \SI{20}{\giga\hertz}, the characteristic radius of the ADW is calculated to be $R_{\text{ADW}} \sim m_a^{-1} \approx O(1)$ cm.

As these objects traverse the Earth, annihilations on their surface produce a flux of mildly relativistic axions with a large intrinsic bandwidth $\Delta\nu/\nu \sim \mathcal{O}(1)$ \cite{Liang_aqn_2020, Budker_AQN_2020}. The resulting broadband spectral shape, shown as the red curve in Fig.~\ref{fig:cdm_aqn_signal_shape}, is fundamentally different from the narrow line expected for CDM axions (in blue). This distribution is derived by mapping the AQN velocity profile from \cite{Fischer_CAST_2018} into the frequency domain via the relativistic relation $\nu = (m_a \gamma)/h$, where $m_a$ is the axion rest mass, effectively broadening the signal as a function of the nugget's Lorentz factor. It is noted that, in contrast to CDM axion searches, relativistic axions do not deposit their energy coherently into a single cavity mode. Hence, in this case, increasing the cavity's quality factor $Q_L$ does not improve sensitivity. The relevant scale is set by the Compton wavelength $\lambda_C = h / (m_a c)$, and the condition \cite{Caspers_daily_2025}:
\begin{equation} 
\label{eq:compton} 
	m_a L \approx \frac{L}{\lambda_C} \gg 1, 
\end{equation} 
which must be satisfied to ensure efficient detection.

In the context of the AQN model, the axion mass is expected to be around $\sim\SI{e-4}{eV}$, corresponding to about \SI{20}{\GHz}. This estimation arises from the typical baryon number of AQNs ($B \sim 10^{25}$). Several indirect astrophysical observations \cite{Sekatchev_2025_glow} fit this mass range, including UV excess in galactic environments, cosmological structure formation, CMB spectral distortions, and IceCube limits on heavy DM interactions \cite{Lawson_2019_trapped}. Therefore, combined with the lattice QCD simulations mentioned above, this mass region is of particular interest for both theoretical and experimental reasons, as it remains largely unexplored by existing haloscope searches.

Interestingly, recent work \cite{Zhitnitsky_mysterious_2025} has connected AQN-induced energy deposition to a wide range of terrestrial and atmospheric anomalies, including stratospheric \cite{Zioutas_2020_stratosphere} and ionospheric disturbances, by identifying AQNs as the source of streaming invisible matter \cite{Bertolucci_2017_sun}. AQNs have also been discussed as a possible contributor to the “Unidentified Falling Objects” (UFO) transient beam losses at the LHC, motivating proposals to reanalyze beam monitor data for potential compact AQN signatures \cite{Zioutas_2024_LHC, Liang_UFO_2026}. These cross-disciplinary connections highlight the relevance and timeliness of dedicated experimental efforts to test the AQN hypothesis through direct axion detection.

\subsubsection{Axion Emission and Daily Modulation}
\label{subsubsec:daily_modulation}

AQNs traveling through the Earth’s atmosphere and interior lose baryon charge due to annihilation processes, which lead to an emission of mildly relativistic ($\beta \sim 0.6$) axions \cite{Fischer_CAST_2018}. In particular, when AQNs traverse the Earth, the estimated averaged energy density within Earth's volume of the corresponding axions is:
\begin{equation}
\rho_a^\text{AQN} \sim 10^{-4} \frac{\langle\Delta B\rangle}{\langle B\rangle} \si[per-mode=symbol]{\giga\electronvolt\per\cubic
\centi\metre} \quad [\text{AQN - related}]
\end{equation}
where $\langle\Delta B\rangle / \langle B\rangle$ denotes the globally averaged fractional baryon number loss during Earth crossing of AQNs derived from full-scale Monte Carlo simulations. This fraction is estimated to be in the range of 10\% to 30\%, with the precise value dependent on the size distribution of the AQNs 
\cite{Liang_aqn_2020}.

\begin{figure}[!htb]
    \centering
    \includegraphics[width=0.5\linewidth]{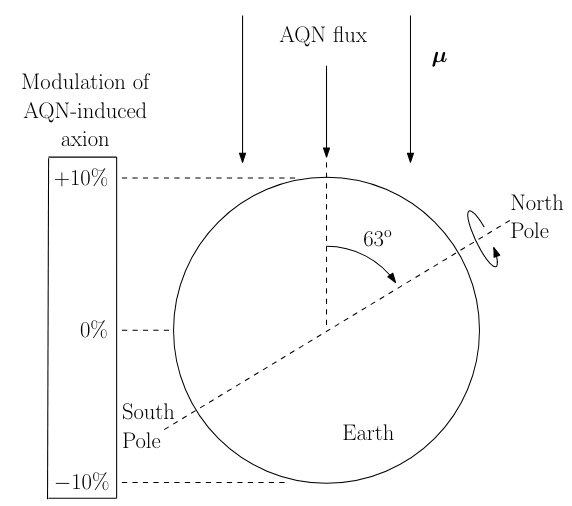}
    \caption{Daily modulation mechanism for AQN-related axions. The \SI{63}{\degree} inclination of the DM wind, originating from the galactic plane relative to the celestial equator, results in approximately 10\% more axions being emitted toward the upper hemisphere than the lower. As the Earth rotates, this anisotropy leads to a detectable $\sim$10\% daily modulation in the axion signal. The modulation amplitude is further enhanced for detectors located at lower latitudes \cite{Liang_aqn_2020}.}
    \label{fig:daily_modulation_mechanism} 
\end{figure}

Therefore, the reduction of AQN size as it crosses the Earth due to annihilation processes within the Earth will produce a distinct signal on axion detectors since the flux of AQN-induced axions is sensitive to the local mass loss of AQNs. This means that more axions are emitted when facing the AQN wind compared to facing away from it (see Fig.~\ref{fig:daily_modulation_mechanism}). Then, as the Earth rotates daily (with respect to the Galactic frame, i.e., a sidereal day of $T_\text{sid}$ =  23h 56m 4s), a daily modulation is expected with a higher amplitude for detectors built in lower latitude regions with respect to the north pole \cite{Liang_aqn_2020, Budker_AQN_2020}.

The daily modulation can be parametrized as:
\begin{equation}
\label{eq:daily_modulation}
A_d(t) \equiv [1 + k_d \cos(\Omega_d t - \phi_0)],
\end{equation}
where $\Omega_d=2\pi / T_\text{sid}$ is the angular frequency of the daily modulation, $\phi_0$ a phase shift, and $k_d$ the relative modulation amplitude in the power spectrum, which is estimated as \cite{Liang_aqn_2020}:
\begin{equation}
\label{eq:aqn_relative_modulation_amplitude}
k_d \approx \frac{1}{3} \cdot \frac{\langle \Delta B \rangle}{\langle B \rangle},
\end{equation}
$A_d(t)$ is considered constant daily, but it gradually changes over time because the DM wind's direction varies with Earth's position and orientation. For example, for opposite seasons like summer vs. winter we would expect a phase shift of $\phi_0 \simeq \pi$.

Previous searches for daily modulation signatures were carried out using the CAST-CAPP \cite{Adair_CAST-CAPP_2022} detector at CERN \cite{Caspers_daily_2025}, which was repurposed to investigate a potential AQN-related axion signal \cite{Fischer_CAST_2018, Liang_aqn_2020}. Although a daily modulation was observed, its phase and amplitude correlated with ambient temperature cycles, implying temperature-induced gain drifts in the amplification chain. While this study provided a valuable proof-of-concept for modulation-based axion searches, the inability to isolate systematics from potential signals limited the actual sensitivity. Nevertheless, CAST-CAPP demonstrated that daily modulation is an experimentally accessible observable and helped identify critical systematic effects for new experiments to overcome.

\subsection{Streaming Dark Matter and Transient Events}
\label{subsec:streaming_dm}

A growing body of theoretical and numerical work suggests that the local DM distribution may contain significant substructures in the form of streams \cite{Vogelsberger_2011_streams} or miniclusters \cite{Tkachev_1991_minicluster, Kolb_1993_miniclusters}, in addition to the standard isotropic halo. These structures are expected to have much smaller velocity dispersion than the standard halo (e.g. $\Delta v_a^\text{stream} \sim 10^{-17}c$) \cite{Kryemadhi_gravitational_2023}, resulting in extremely narrow spectral features \cite{OHare_2023_Minicluster, OHare_2025_finegrained}, with linewidths much smaller compared to the typical \si{\kHz} width for virialized halo axions \cite{Knirck_2018_directional}.

Gravitational focusing by solar system bodies such as the Sun, Earth, and Moon can occasionally amplify the local density of low-velocity ($v\sim 10^{-3}c$) streaming DM towards the Earth by several orders of magnitude. This is because for slow-moving DM particles, the gravitational deflection increases with $1/\text{velocity}^2$. As an example, under optimal alignment, the flux amplification due to the Sun can reach values up to $10^{11}$ for durations ranging from seconds to several hours, depending on the stream velocity \cite{Kryemadhi_gravitational_2023}. The Moon can also focus DM particles with velocities $\sim 10^{-4}c$ onto the Earth with a flux amplification of about $10^4$ \cite{Zioutas_2017_search}. Finally, the inner Earth can also act as a high-efficiency gravitational collector for particles with an initial velocity of about \SI[per-mode=symbol]{17}{\km\per\second} with respect to the geo-centre with a flux increase of about $10^9$ \cite{Sofue_2020_focusing}.

Standard axion haloscope searches average data over long periods and use frequency-domain analyses optimized for the expected \si{kHz}-scale width of the standard halo axion model. This approach effectively averages out short-duration, narrow-band signals from streams or miniclusters, especially when the resolution bandwidth is not sufficiently small. As a result, transient or streaming events may go unnoticed in conventional analyses.

To address this, a dedicated search strategy based on high cadence spectral analysis is required to detect sharp spikes and transient features. By processing data in short time intervals, short-lived power excesses can be identified while maintaining the frequency resolution required to resolve narrow DM substructures. Since the timing of such events is unpredictable, continuous operation is essential. Observational hints suggest an increased rate of anomalies during December–January, based on unexplained disturbances in the ionosphere \cite{Bertolucci_2017_sun} and stratosphere \cite{Zioutas_2020_stratosphere}. This period coincides with the Earth–Sun–Galactic Center alignment, which may enhance DM stream flux via gravitational focusing. Such an approach offers a unique window into the fine structure of DM and the early-universe origin of axion DM.

\section{Experimental Design}
\label{sec:experimental_design}

\subsection{The Thin-Shell ADAMOS Cavity} 
\label{subsec:thin_shell}

The ADAMOS experiment aims to pioneer a new class of fixed-frequency haloscope searches for axions in the high-frequency regime around \SI{20}{\GHz} corresponding to an axion mass of approximately \SI{82.7}{\micro\eV}. Designed to operate inside an existing \SI{14}{\tesla} warm-bore solenoid magnet at Universität Hamburg, ADAMOS addresses key limitations encountered in past axion searches, like: the inability to probe high axion masses due to volume scaling, the need for daily modulation sensitivity, and the capability for transient event detection.

\begin{figure}[htb!]
\centering
\includegraphics[width=0.32\textwidth]{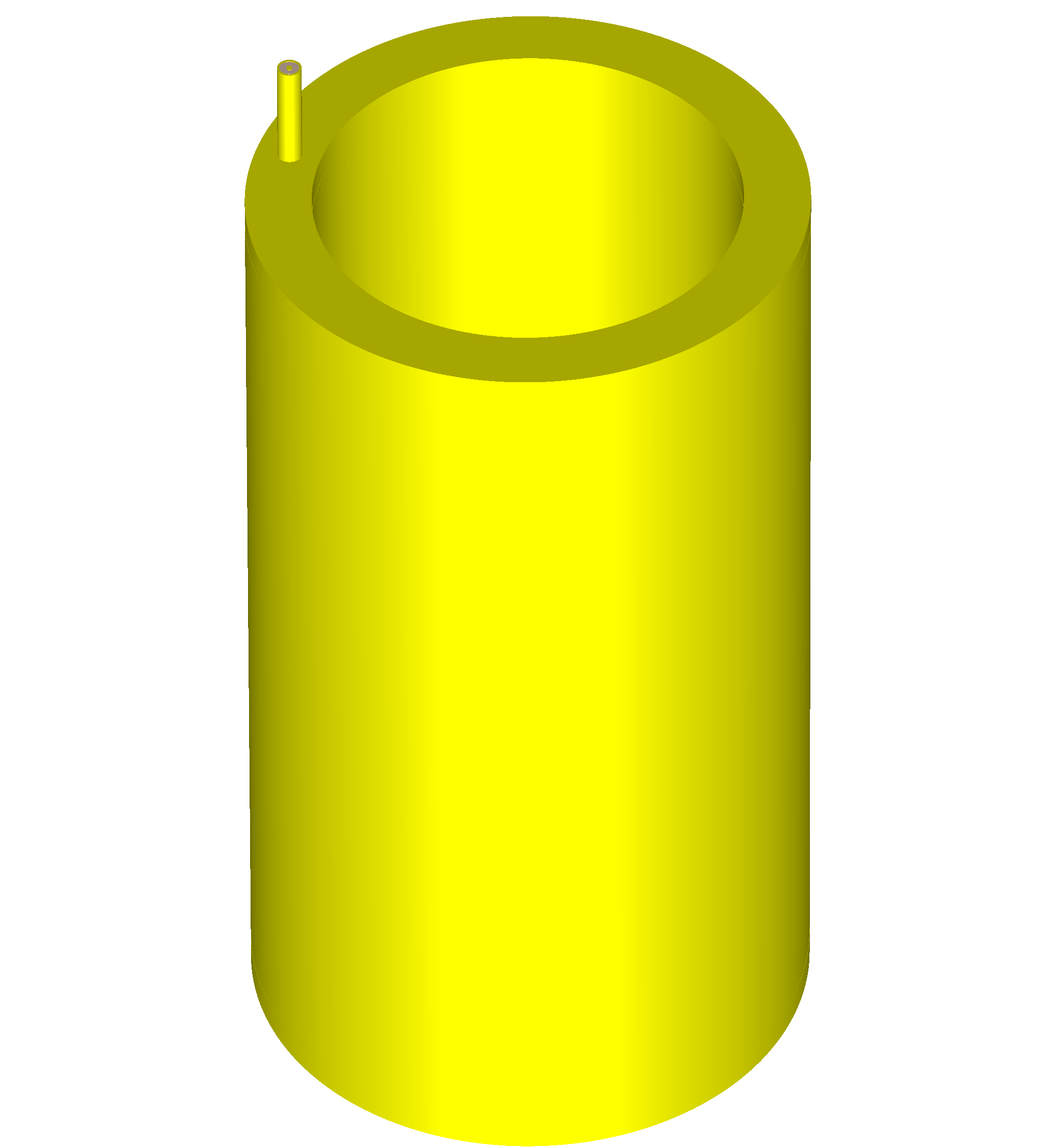}
\includegraphics[width=0.38\textwidth]{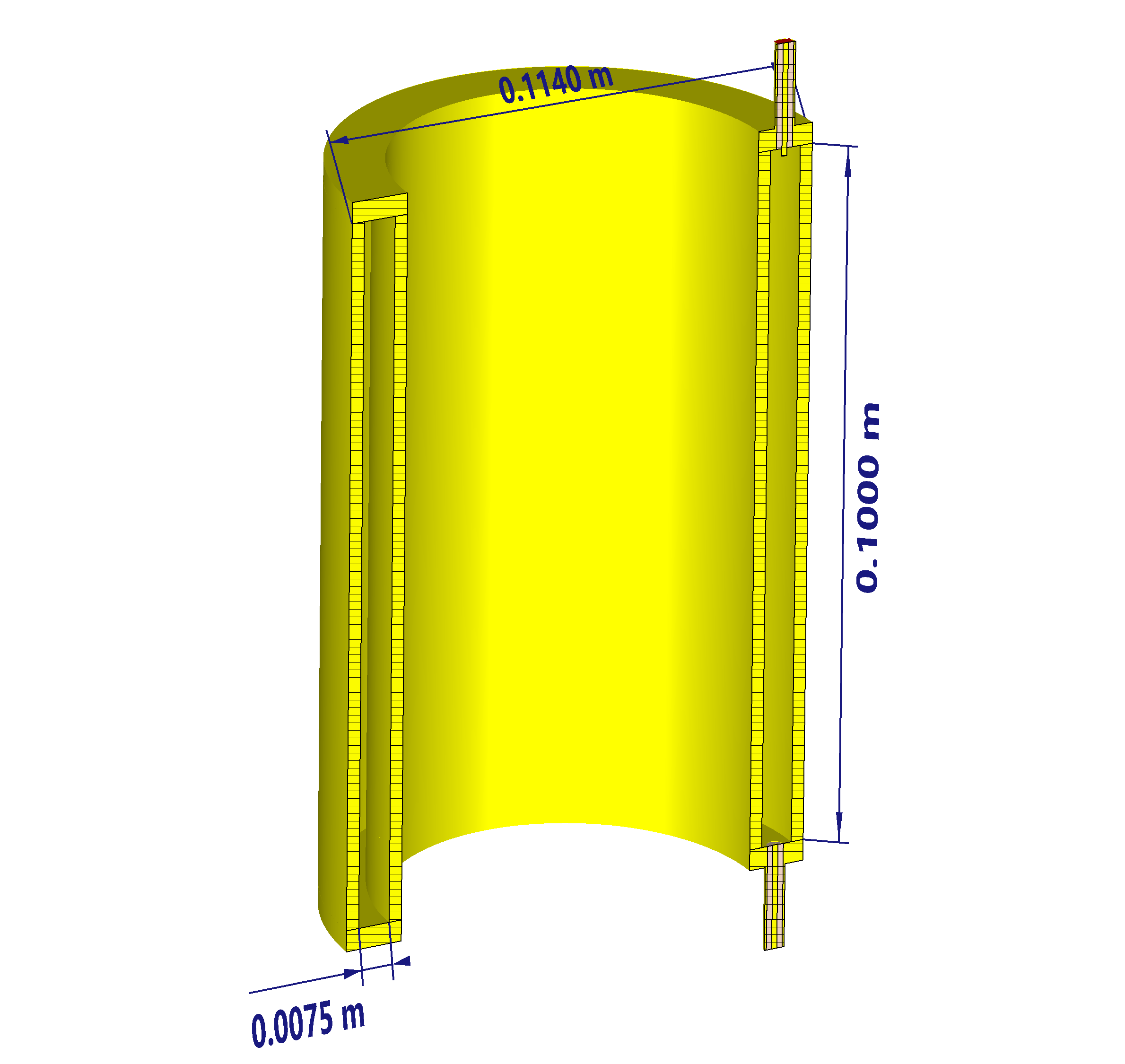}
\caption{3D model of the ADAMOS cavity. The thin-shell geometry consists of two concentric cylinders with a uniform \SI{7.5}{\mm} gap, which defines the annular volume where the pseudo-TM$_{010}$ mode is supported. The CAD rendering is truncated for visualization; the full experimental cavity has a length of \SI{400}{\mm} and an outer diameter of \SI{125}{\mm}.}
\label{fig:adamos_design}
\end{figure}

ADAMOS utilizes a novel “thin-shell” cavity geometry \cite{Kuo_2019_large}, inspired by recent implementations \cite{DiVora_2025_axion}. This design, shown in Fig.~\ref{fig:adamos_design}, decouples resonant frequency from cavity volume, overcoming a key limitation of traditional haloscopes at high frequencies. The cavity consists of two concentric OFHC cylinders, each \SI{8}{\mm} thick. The outer cylinder has a \SI{125}{\mm} outer diameter, matching the \SI{14}{\tesla} magnet's warm-bore, while the inner cylinder creates a \SI{7.5}{\mm} gap with a \SI{94}{\mm} outer diameter. The cavity has a total length of \SI{400}{\mm}, yielding a cylindrical annular volume of \SI{0.96}{\litre}. This is about $25$ times the detection volume of a typical cylindrical cavity of the same length operating at the same frequency ($V \sim \SI{0.04}{\litre}$). This geometry supports a pseudo-TM$_{010}$ mode at the target frequency of \SI{20}{\GHz}, aligning with theoretical predictions for axion masses arising from the AQN DM model and supported by QCD lattice calculations.

\begin{figure}[htb!]
\centering
\includegraphics[width=0.485\textwidth,trim={0 0 10cm 1cm},clip]{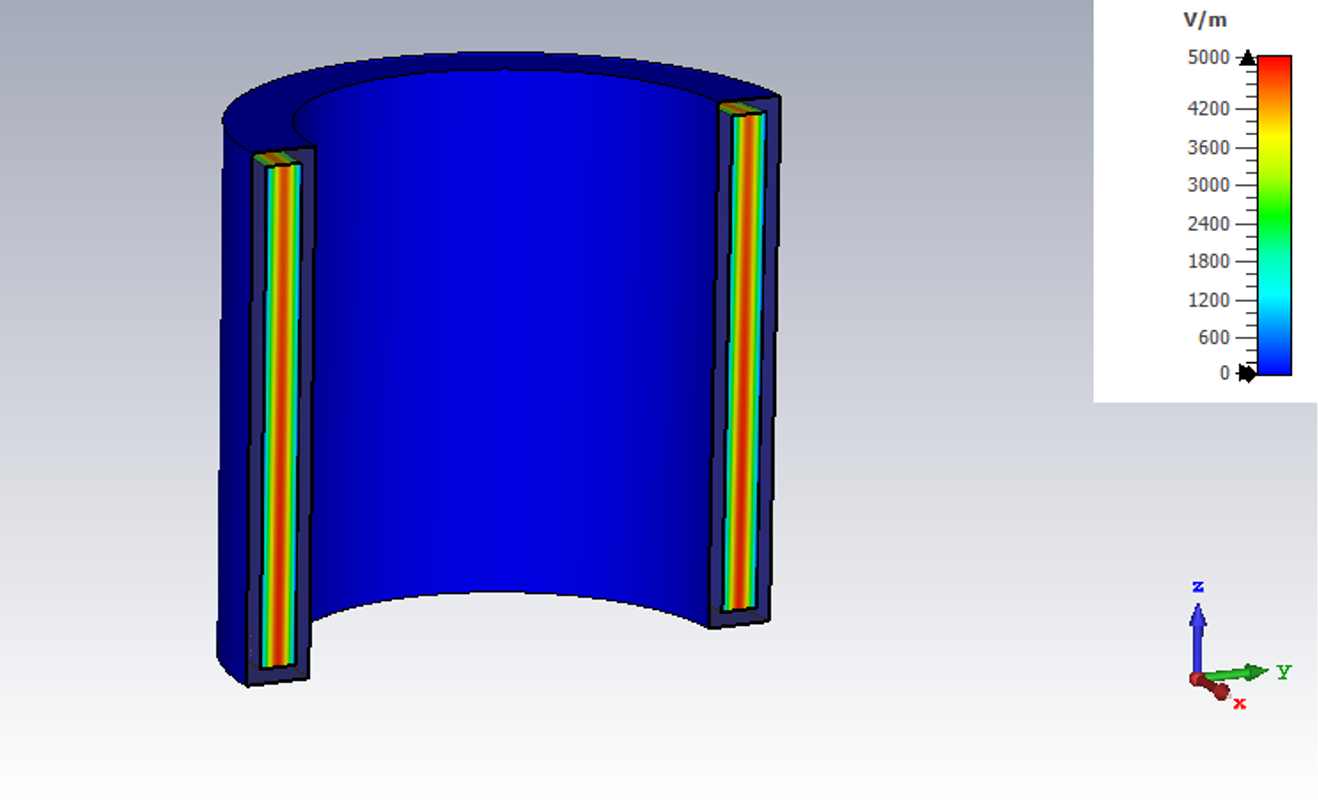}
\includegraphics[width=0.49\textwidth,trim={0 0 10cm 1cm},clip]{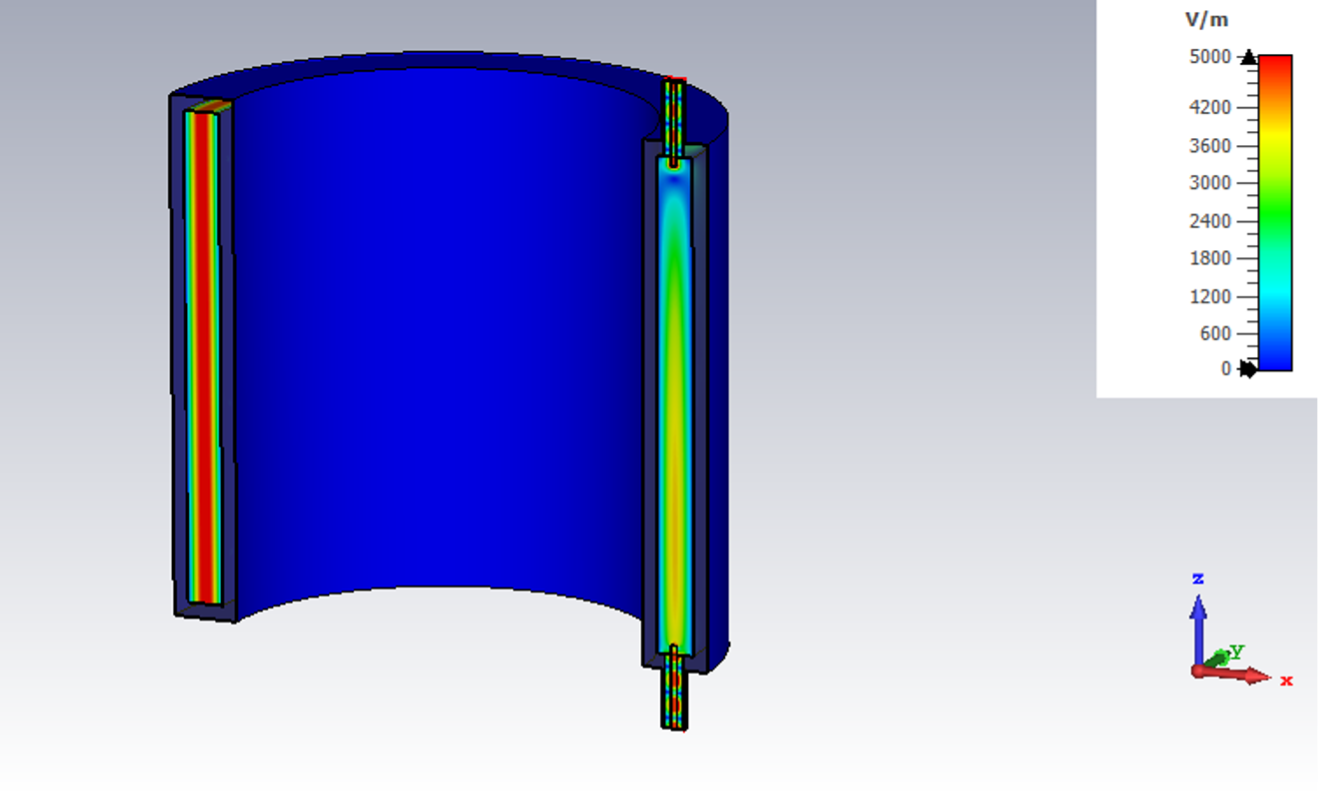}
\caption{Simulation of the distribution of the fundamental pseudo-TM$_\text{010}$ mode inside the empty ADAMOS cavity \textbf{(left)} and with the two input and output antennas \textbf{(right)} using CST Studio Suite. The color scale indicates the normalized electric field magnitude, with red representing the maximum field intensity.}
\label{fig:adamos_simulation}
\end{figure}

To validate the electromagnetic performance, a combined eigenmode and frequency-domain study was performed using CST Studio Suite with adaptive mesh refinement. Numerical convergence was ensured by iteratively refining a 2-million-element mesh until the relative variation in resonant frequency fell below $10^{-4}$. Eigenmode analysis of the ideal, empty cavity yields a form factor $C = 0.815$, compatible with previous estimations \cite{Kuo_2019_large}. When the physical antennas are introduced, as shown in Fig.~\ref{fig:adamos_simulation}, the fundamental pseudo-TM$_{010}$ mode is identified at $f_0 = \SI{19.95}{\GHz}$ with a slightly reduced form factor $C \approx 0.790$. This confirms a strong overlap between the cavity mode and the externally-applied static uniform magnetic field $B_0$, despite the high-frequency operation and the coupling-port perturbations.

\begin{figure}[h!]
\centering
\includegraphics[width=0.75\textwidth]{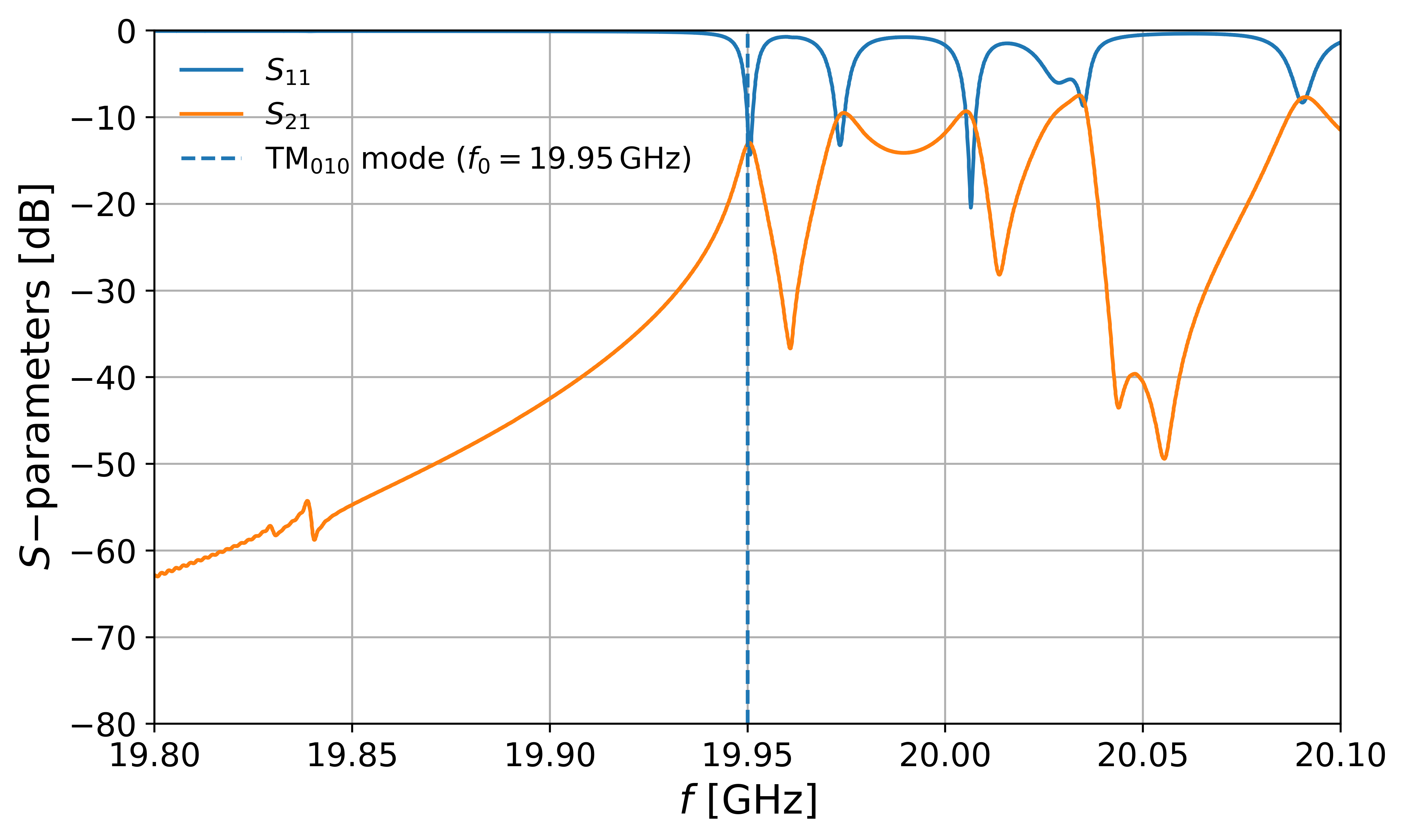}
\caption{Simulated scattering parameters ($S_{11}$ and $S_{21}$) for the ADAMOS thin-shell cavity. The weak-coupling input (\SI{0.5}{\mm}) and output (\SI{1.1}{\mm}) antennas are optimized for near-critical coupling ($\beta_\text{out} \approx 1.0$), yielding $Q_L = 4100$ and a \SI{3}{\dB} bandwidth of \SI{4.9}{\MHz} at \SI{19.95}{\GHz}.}
\label{fig:adamos_sparams}
\end{figure}

The cavity's resonance parameters were further extracted from frequency-domain simulations of the scattering parameters, shown in Fig.~\ref{fig:adamos_sparams}. The system employs a two-port configuration with an input antenna pin length of about \SI{0.5}{\mm} (weak coupling) and an output antenna pin length of about \SI{1.1}{\mm} (near-critical coupling). A simultaneous complex fit of $S_{11}$ and $S_{21}$ was used to determine the coupling coefficients, yielding $\beta_\text{in} \approx 0.09$ and $\beta_\text{out} \approx 1.0$. The extracted parameters yield a loaded quality factor of $Q_L = 4100$, corresponding to a \SI{3}{\dB} resonance bandwidth of $\Delta f = f_0 / Q_L \approx \SI{4.9}{\MHz}$. This is consistent with expectations for OFHC copper at room temperature.

This configuration maintains a liter-scale detection volume while achieving an electromagnetic figure of merit $C Q_L \approx 2.8 \times 10^3$ at \SI{20}{\GHz}. Finally, from Eq.~\ref{eq:compton}, given ADAMOS's dimensions, $L m_a \approx 1.7 \times 10^2 \gg 1$, satisfying the requirement for relativistic AQN-induced axions.

\subsection{RF Chain and Readout}
\label{subsec:rf_chain}

\begin{figure*}[htb!]
\centering
\includegraphics[width=0.95\textwidth]{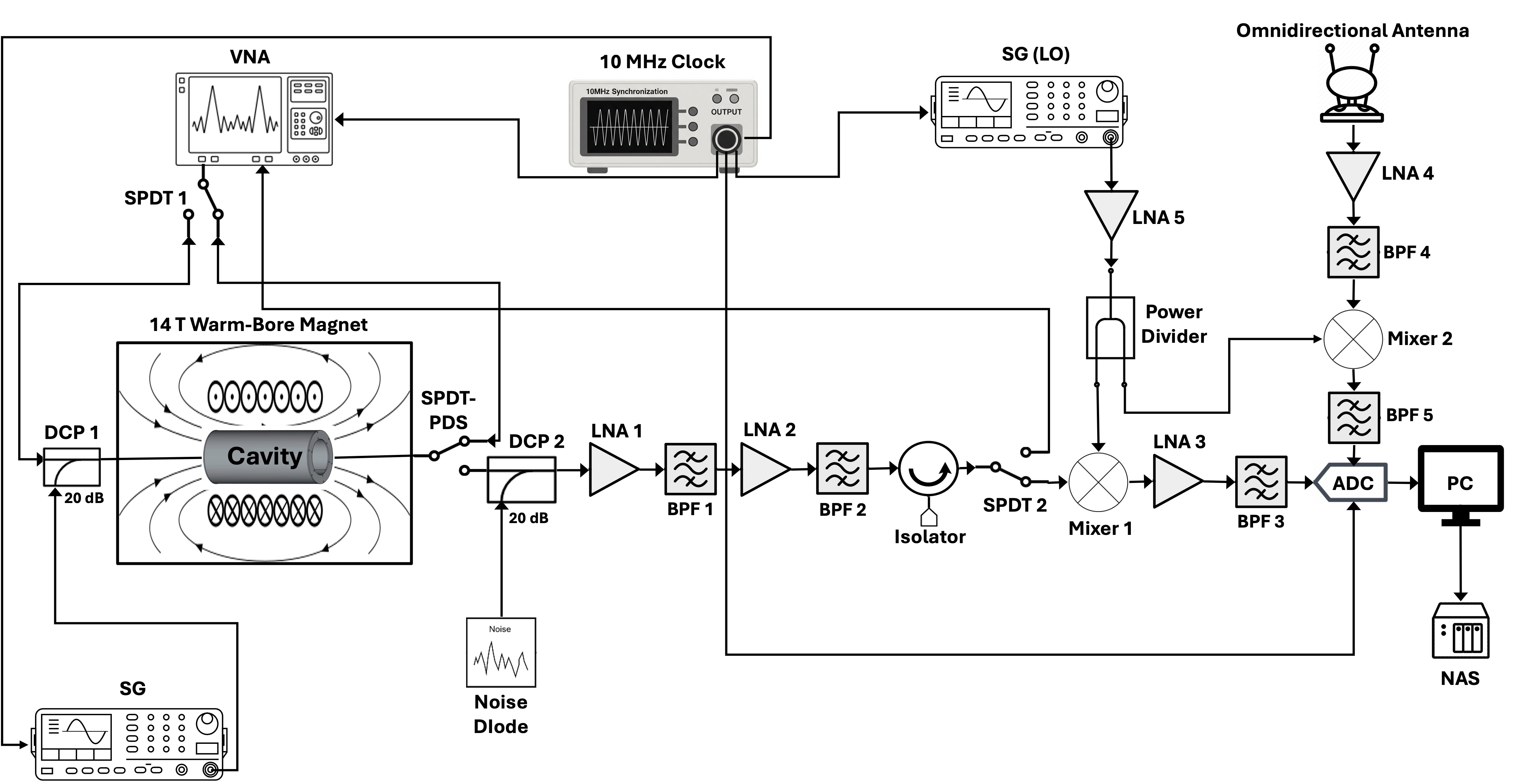}
\caption{Schematic diagram of the proposed ADAMOS experimental setup with the arrows indicating the signal direction. The various acronyms correspond to: Directional Coupler (DCP), Low Noise Amplifier (LNA), Signal Generator (SG), Vector Network Analyzer (VNA), Single-port-double-through (SPDT) Pin-diode switch (PDS), Band-pass filter (BPF), Low-pass Filter (LPF), Analog-to-Digital Converter (ADC), Local Oscillator (LO).}
\label{fig:adamos_schematic}
\end{figure*}

The signal from the cavity is processed through a sophisticated RF chain designed for high sensitivity and robust calibration, as detailed in the schematic of Fig.~\ref{fig:adamos_schematic}. The output from the cavity is first routed through a non-magnetic, single-pole double-throw (SPDT) pin-diode switch (PDS). This switch allows the system to alternate between a transmission measurement mode for data acquisition and a reflection measurement mode for characterizing the cavity's resonant frequency and coupling.

Signal routing is managed by mechanical single-pole double-throw (SPDT) switches. SPDT 1 selects between the cavity input and output ports for reflection coefficient ($S_{11}$) measurements to characterize antenna coupling, while SPDT 2 routes the cavity output either to the down-conversion chain for data acquisition or directly to a vector network analyzer (VNA) for transmission ($S_{21}$) measurements. The antenna coupling will remain fixed during operation, with periodic $S_{11}$ checks used to monitor any drift in the coupling.

The signal from the cavity is then amplified by a cascade of two low-noise amplifiers (LNAs), LNA 1 and LNA 2, with band-pass filters (BPFs) placed after each amplifier stage to suppress out-of-band noise. An isolator ensures amplifier stability by suppressing reflections back through the signal chain. ADAMOS will use a heterodyne receiver scheme by down-mixing the amplified $\sim$\SI{20}{\GHz} signal to an intermediate frequency (IF) of \SI{10}{\MHz} using a balanced mixer and a local oscillator (LO) amplified by LNA 5. This down-conversion reduces the required sampling rate and simplifies subsequent digitization and processing. The IF signal is then amplified (LNA 3) and passed through BPF 3 to define the analysis bandwidth and provide anti-aliasing protection. The conditioned signal is finally digitized by a 16-bit analog-to-digital converter (ADC) with a sampling rate of up to \SI[per-mode=symbol]{500}{\mega\sample\per\second}. To maintain phase coherence, the LOs, VNA, and ADC are all synchronized to a common \SI{10}{\mega\hertz} laboratory clock (Fig.~\ref{fig:adamos_schematic}). For the narrow-line CDM search, the clock's stability is more than an order of magnitude higher than the required coherence time of $\approx \SI{30}{\micro\second}$, preventing spectral smearing. Furthermore, the common reference ensures that transient events can be time-correlated between the cavity and veto channels with sub-microsecond precision. Phase noise in the IF chain is negligible for these power-based measurements.

A parallel readout chain processes signals from the EMI/EMC veto system (see Sect.~\ref{subsec:calib_veto} for more details). The omnidirectional antenna output is amplified by LNA 4, filtered by BPF 4, and down-mixed to the \SI{10}{\MHz} IF with the same mixer configuration that is also used for the cavity signal. The down-converted veto signal then passes through BPF 5 for bandwidth definition and anti-aliasing before digitization by a second ADC channel. This identical processing ensures synchronous acquisition and direct comparability between the cavity and veto channels.


The data acquisition (DAQ) system is based on a newly developed framework that is designed for real-time and high-resolution data processing \cite{Cezar_2026_modular}. This includes the synchronous digitization and Fourier transformation of the two down-mixed IF signals (cavity and veto). Within the \SI{125}{\mega\hertz} analysis bandwidth, defined by the single-channel ADC sampling rate, the DAQ performs a multi-resolution analysis: For the transient search, it computes high-resolution power spectral densities over $\sim$\SI{1}{\minute} intervals, providing the temporal and spectral resolution (${\sim}\SI{16.7}{\milli\hertz}$) necessary to resolve narrow DM substructures. Simultaneously, for the CDM search, the captured time-domain signal is divided into multiple overlapping segments. An FFT is applied to each segment to generate a set of spectra, each with a frequency resolution of at least \SI{3.6}{\kilo\hertz} and an effective temporal resolution reaching values as low as $\SI{278}{\micro\s}$. This provides at least 10 bins across the predicted \SI{36}{\kilo\hertz} FWHM of a virialized halo at \SI{20}{\GHz} (see Fig.~\ref{fig:cdm_aqn_signal_shape}), ensuring the spectral shape can be precisely characterized, while also allowing for temporal resolution. Finally, the daily modulation analysis utilizes the integrated power of these spectral sets, tracked over sidereal time, to identify periodic signal variations. This integrated architecture allows ADAMOS to pursue all three primary science goals simultaneously.

\subsection{Calibration and Veto Systems}
\label{subsec:calib_veto}

The ADAMOS design incorporates comprehensive calibration and monitoring systems to control systematic effects that have limited previous axion searches, particularly for daily modulation signatures \cite{Caspers_daily_2025}.

Calibration is performed using directional couplers strategically placed in the RF chain. DCP~1, placed at the cavity input, injects a calibrated pilot tone from a signal generator (SG) into the main signal path for gain stability monitoring. DCP~2, placed before LNA~1, periodically injects a broadband noise signal from a noise diode with a known Excess Noise Ratio (ENR) for Y-factor noise‐temperature measurements of the entire amplification chain. The system noise temperature $T_{\text{sys}} \approx \SI{480}{\kelvin}$ is conservatively estimated for room-temperature operation, accounting for the first-stage LNA~1, and any IF-chain noise contributions. Future cryogenic cooling could reduce $T_{\text{sys}}$ by more than $\mathcal{O}(1)$. In addition to the gain calibration, ADAMOS will monitor the TM$_{010}$ resonant frequency through periodic $S_{21}$ transmission measurements and verify both input and output antenna couplings through $S_{11}$ reflection measurements, both carried out using SPDT~1, the PDS switch, and the VNA.

A central innovation is the implementation of an automated, frequent calibration cycle operating at approximately one-minute intervals, optimized to track and correct slow thermal drifts while maintaining a high data-taking duty cycle. These measurements provide the necessary data for applying gain and noise corrections during offline analysis, ensuring that any observed power variations can be confidently attributed to astrophysical sources rather than instrumental drifts (see Sect.~\ref{subsec:modulation_sensitivity}).

Complementing the electronic calibration, a multi-layered thermal management strategy will be implemented to ensure environmental stability. The cavity will be thermally isolated from the magnet bore using low-conductivity materials to minimize heat transfer from the cryogenic system. The experimental room will be maintained at a stable temperature through active climate control, and critical RF components, including LNAs and mixers, will be equipped with passive cooling solutions. Comprehensive temperature monitoring using PT100 sensors will provide continuous logging of all major components, creating a detailed thermal profile for correlation with any remaining observed signal variations.

To reject environmental radio-frequency interference, the dedicated EMI/EMC veto system employs an omnidirectional antenna with an identical readout chain to the main cavity channel, as described in Sect.~\ref{subsec:rf_chain}. This system continuously monitors the ambient electromagnetic spectrum around the search frequency of \SI{20}{\GHz}. Signals appearing simultaneously in both the cavity and veto channels are automatically flagged and excluded from analysis as external interferences \cite{Adair_CAST-CAPP_2022}.

\section{Sensitivity Projections}
\label{sec:sensitivity}

The sensitivity of the ADAMOS experiment is evaluated for each of its three primary search channels. The projected performance demonstrates the experiment's capability to probe significant new parameter space for axion dark matter across complementary detection paradigms.

\subsection{Cold Dark Matter Axion Search}
\label{subsec:cdm_sensitivity}

\begin{figure}[htb!]
\centering
\includegraphics[width=0.95\columnwidth]{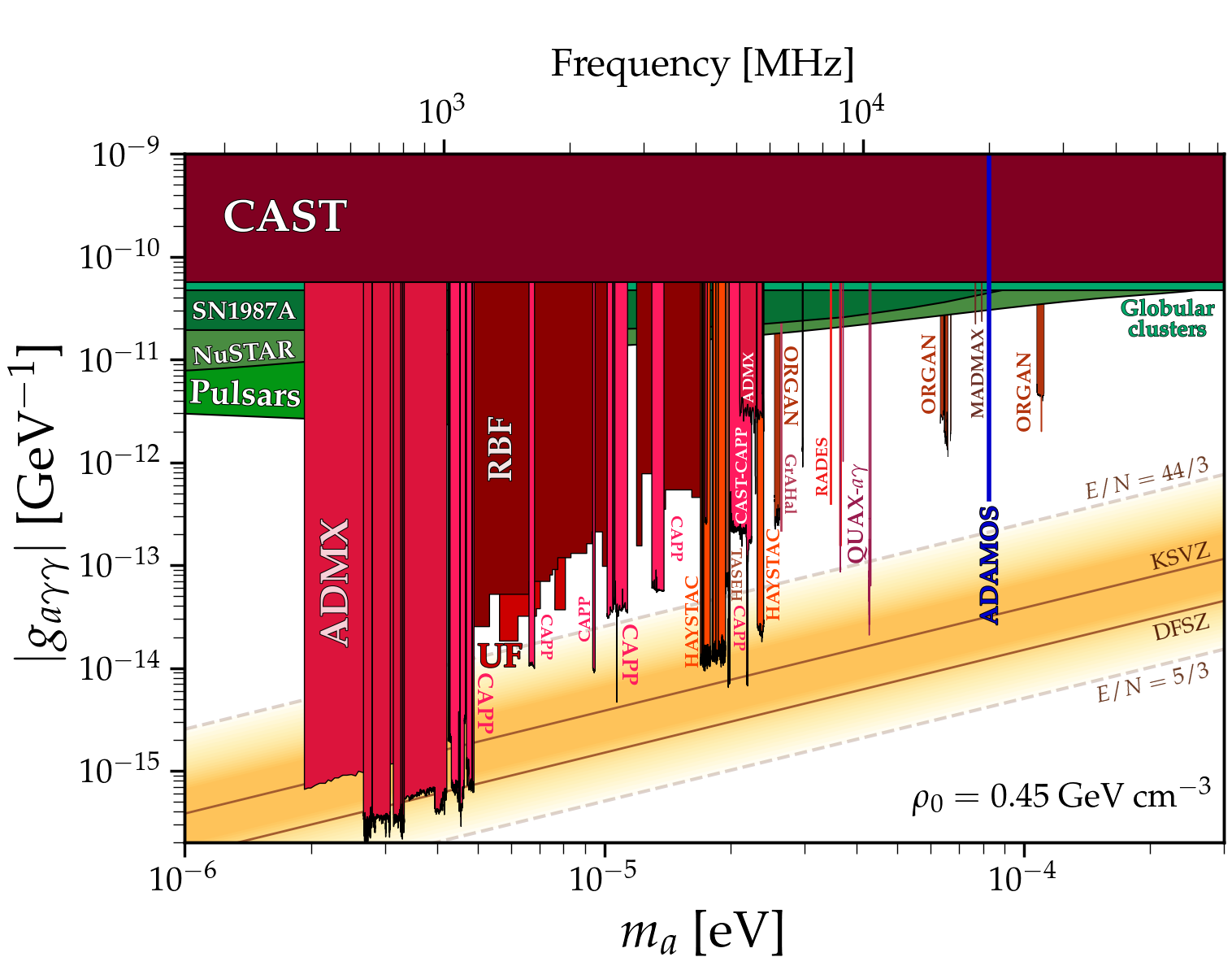}
\caption{ADAMOS projected sensitivity (\textcolor{blue}{in blue}) for 30 days integration for CDM axions at \SI{19.95}{\GHz} compared to existing limits from other haloscopes (in \textcolor{red}{red}) and astrophysics (in \textcolor{green}{green}). The theoretical QCD axion band is also shown (in \textcolor{orange}{orange}). Plot adapted from \cite{OHare_2020_axionlimits}.}
\label{fig:adamos_cdm_sensitivity}
\end{figure}

For the conventional CDM axion search, the expected sensitivity to the axion-photon coupling $g_{a\gamma\gamma}$ can be estimated using the standard haloscope figure of merit \cite{rades_2022_design}:
\begin{equation}
\label{eq:coupling_calculation}
g_{a\gamma\gamma} = \left( \frac{\mathrm{SNR} k_B T_{\mathrm{sys}}}{\rho_a B_0^2 V C \kappa Q_l}\right)^{1/2} \left( \frac{m_a}{Q_a t} \right)^{1/4},
\end{equation}
where $k_B$ is the Boltzmann constant, and $\kappa = \frac{1+\beta_\text{out}}{\beta_\text{out}}$ is the coupling efficiency. The rest of the parameters and their values for ADAMOS are summarized in Tab.~\ref{tab:adamos_cdm_parameters}. For an integration time of 30 days at the fixed frequency of \SI{19.95}{\GHz}, ADAMOS is projected to reach a sensitivity of $g_{a\gamma\gamma} = \SI{6.88e-13}{\GeV^{-1}}$ for an axion mass of \SI{82.51}{\micro\eV}. This represents a significant improvement over existing limits in this frequency range, as shown in Fig.~\ref{fig:adamos_cdm_sensitivity}. 

\begin{table}[htb!]
\centering
\caption{Numerical parameters used in ADAMOS for CDM axion sensitivity estimates.}
\begin{tabular}{|l|l|l|}
\hline
\label{tab:adamos_cdm_parameters}
\textbf{Symbol} & \textbf{Description} & \textbf{Value} \\
\hline
$f_0$ & Resonant frequency & \SI{19.95}{\GHz} \\
$m_a$ & Axion mass & \SI{82.51}{\micro\eV} \\
$\mathrm{SNR}$ & Target Signal-to-noise ratio & 5 \\
$T_{\mathrm{sys}}$ & System noise temperature & \SI{480}{\kelvin} \\
$B_0$ & Magnetic field strength & \SI{14}{\tesla} \\
$V$ & Cavity volume & \SI{0.96}{\litre} \\
$C$ & TM$_{010}$ form factor & 0.79 \\
$\rho_a$ & Local axion density & \SI[per-mode=symbol]{0.45}{\GeV\per\cubic\cm} \\
$\beta_\text{out}$ & Output antenna coupling & 1 \\
$Q_a$ & Axion quality factor & $10^6$ \\
$Q_L$ & Loaded cavity quality factor & 4100 \\
$t$ & Integration time & 30 days \\
\hline
\end{tabular}
\label{tab:adamos_numeric_values}
\end{table}

In addition to the primary frequency-domain CDM search, the long-term operation of ADAMOS provides a secondary sensitivity gain through the conventional daily modulation expected from the Earth's rotation with respect to the Galactic DM wind \cite{Ling_2004_diurnal, Gramolin_2022_spectral}. Although this modulation is small (of order $10^{-3}$), continuous operation over multi-year timescales enables the accumulated power measurements to be coherently folded over the sidereal period. This effectively increases the signal-to-noise ratio for any temporally modulated component, offering a parasitic improvement in sensitivity to $g_{a\gamma\gamma}$ beyond what is achieved from the frequency-domain search alone. While this effect is subdominant compared to the primary haloscope sensitivity, it provides an additional and independent consistency check for any putative CDM axion signal.

\subsection{AQN Daily Modulation Search}
\label{subsec:modulation_sensitivity}

The sensitivity to daily modulation signatures from Axion Quark Nuggets follows a different scaling than the conventional haloscope search. For relativistic axions with large bandwidth ($\Delta\nu/\nu \sim 1$), the coherence is lost, resulting in quality factors of $Q_a^\text{AQN} = Q_L^\text{AQN} = 1$. In this regime the cavity does not provide resonant enhancement, and the sensitivity scales with the total collected power rather than $Q_L$.

As described in Sect.~\ref{subsubsec:daily_modulation}, the expected signal manifests as a time-varying excess in the integrated power output of the cavity, modulated with sidereal-day periodicity due to Earth's rotation through the galactic AQN wind. A seasonal comparison will also search for the predicted $\pi$ phase shift in the modulation phase $\phi_0$ from Eq.~\ref{eq:daily_modulation}. The amplitude of this daily modulation depends on geographic latitude due to the inclination between Earth's rotational axis and the incoming axion flux. From the azimuthal variation of axion flux $P_a (\theta)$ shown in Fig.~1 of Ref.~\cite{Caspers_daily_2025}, the expected daily modulation amplitude in the signal power at the ADAMOS site in Hamburg (latitude $\approx 53.5^\circ$, corresponding to $\cos\theta = 0.5937$) can be estimated to $5.4\%$. To detect such a modulation at $5\sigma$ significance, the residual gain drift after calibration must remain well below the statistical noise level. Assuming the one-minute calibration cycles described in Sect.~\ref{subsec:calib_veto}, the allowable residual fractional gain variation is required to be $\mathcal{O}(10^{-3})$ over hourly timescales. This is achievable with continuous pilot-tone monitoring and frequent Y-factor recalibration, ensuring that instrumental fluctuations do not mask the astrophysical signal.

A first search for periodic signatures in the calibrated power stream will be performed using a Lomb–Scargle periodogram. This provides a model-independent check for excess power near the sidereal frequency and serves as an initial indicator for daily modulation. The statistical significance of any detected signal is evaluated against the null (white-noise) hypothesis by computing the False-Alarm Probability (FAP). This calculation follows the standard prescription for unevenly sampled data, explicitly accounting for the number of independent frequencies searched to maintain a robust statistical threshold.

For a more sensitive and phase-coherent test, particularly over long data sets, we will employ an epoch-folding analysis, a standard technique in periodic signal searches. Each calibrated and gain-corrected (see Sect.~\ref{subsec:calib_veto}) power measurement $P_i$ at time $t_i$ will be mapped to the expected sidereal phase:
\begin{equation}
\phi_i = 2\pi \left( \frac{t_i - t_0}{T_{\mathrm{sid}}} \right) \bmod 2\pi ,
\end{equation}
where $t_0$ is a reference epoch. The data will then be grouped by phase and averaged to form the folded profile:
\begin{equation}
\langle P(\phi) \rangle =
\frac{1}{N_\phi}
\sum_{i \in \phi\text{-bin}} P_i ,
\end{equation}
which coherently stacks all measurements at the same phase across days or months. A true sidereal modulation increases in significance approximately as $\sqrt{N_{\mathrm{days}}}$, while uncorrelated noise will average down.

To maintain coherence for long-term integrations, the sidereal phase assignment must account for the relative orientation between Earth's rotation and the galactic AQN wind:
\begin{equation}
\phi_i = 2\pi \left( \frac{t_i - t_0}{T_{\text{sid}}} \right) - \Phi_{\text{AQN}}(t_i) \quad \bmod 2\pi,
\end{equation}
where $\Phi_{\text{AQN}}(t)$ is determined by the AQN model, which predicts the phase relationship between Earth's rotation and the incoming flux. Crucially, the model predicts a $\pi$ phase shift between summer and winter observations. 

The final folded light curve will then be tested against the null (flat) hypothesis using a $\chi^2$ statistic with $N_\phi - 1$ degrees of freedom, or a likelihood-ratio test, and can be fitted with Eq.~\ref{eq:daily_modulation} to extract the modulation amplitude and phase. For a sinusoidal daily modulation, the detection significance will scale with the square root of the total observation time, 
\begin{equation} 
\mathrm{SNR}_{\mathrm{mod}} \propto k_d \sqrt{N_{\mathrm{days}}},
\end{equation}
since the modulation adds coherently after folding, while the noise averages down as $1/\sqrt{N_{\mathrm{days}}}$. Long-term, phase-coherent integration is therefore essential for maximizing sensitivity. A detection claim will then be defined by a post-trial significance exceeding 5$\sigma$, while 3$\sigma$ will be used for exclusion limits, ensuring robustness against look-elsewhere effects and long-term systematic drifts.

\subsection{Transient Event Sensitivity}
\label{subsec:transient_sensitivity}

For transient signals from DM streams, the sensitivity improves dramatically during flux enhancement events \cite{Kryemadhi_gravitational_2023}. During such events, the local density $\rho_a$ can be boosted by several orders of magnitude above the nominal value $\rho_a = \rho_\text{halo} = \SI[per-mode=symbol]{0.45}{\GeV\per\cubic\cm}$. Since the minimum detectable axion-photon coupling scales as $g_{a\gamma\gamma} \propto 1/\sqrt{\rho_a}$, as seen in Eq.~\ref{eq:coupling_calculation}, possible flux enhancements can improve the experimental reach. 

\begin{figure}[htb!]
\centering
\includegraphics[width=0.8\columnwidth]{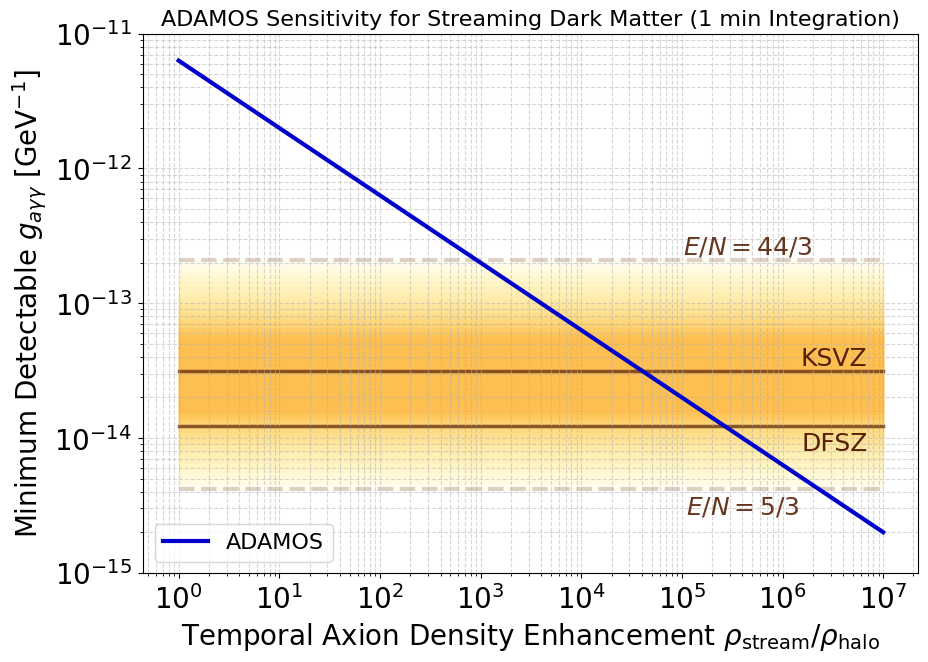}
\caption{Minimum detectable axion-photon coupling $g_{a\gamma\gamma}$ for the ADAMOS experiment as a function of the local axion density enhancement $\rho_{\mathrm stream}/\rho_{\mathrm halo}$, assuming a fixed integration time of 1 minute. The $x$-axis represents the enhancement of the local axion density due to a possible DM stream through gravitational focusing by the solar system bodies.}
\label{fig:adamos_stream_sensitivity}
\end{figure}

Fig.~\ref{fig:adamos_stream_sensitivity} shows how the ADAMOS sensitivity improves as a function of the local axion density enhancement, covering the range from the standard halo value up to $10^7$ times higher density. For a high-resolution 1-minute integration at \SI{19.95}{\GHz}, the required density enhancement to reach the theoretical Kim-Shifman-Vainshtein-Zakharov (KSVZ) axion model \cite{KSVZ1,KSVZ2} is about $\sim 9.8 \times 10^4$, while for the Dine-Fischler-Srednicki-Zhitnitsky (DFSZ) axion model \cite{DFSZ1, DFSZ2} it is $ \sim 6.4 \times 10^5$. These enhancement factors are well within the range expected from gravitational focusing, meaning that during such streaming events, ADAMOS could probe well-motivated QCD axion parameter space, which is otherwise inaccessible with standard halo assumptions.

\section{Conclusion and Outlook}
\label{sect:conclusions}

The ADAMOS experiment explores a novel approach to high-frequency ($\sim$\SI{20}{\GHz}) axion DM detection, targeting a theoretically motivated mass range predicted by both lattice QCD calculations \cite{Borsanyi_2016_mass} and AQN models \cite{Zhitnitsky_2003_aqn, Liang_aqn_2020, Budker_AQN_2020}. By unifying three distinct scientific targets within a single multi-channel haloscope platform, ADAMOS overcomes key limitations of previous searches \cite{Caspers_daily_2025}. The experimental design is built on three central technical innovations:
\begin{itemize}
    \item \textbf{Thin-Shell Cavity Geometry:} A thin-shell design, comprised of two concentric cylinders, preserves a large detection volume of \SI{0.96}{\litre} around \SI{20}{\GHz}, overcoming practical limitations and the severe volume loss inherent to conventional cylindrical cavities at this frequency.
    \item \textbf{Quasi-continuous Calibration:} An automated $\sim$\SI{1}{\minute} calibration cycle, combining noise-diode Y-factor measurements and a pilot-tone system, corrects any temperature-dependent gain and noise drifts, which is essential for daily modulation and transient-event searches.
    \item \textbf{Dedicated multi-layered, multi-channel and real-time DAQ system:} A high-duty cycle, multi-channel RF and digitization chain enables real-time processing without hardware reconfiguration for the different axion searches. ADAMOS is the first haloscope explicitly designed for concurrent, high-sensitivity searches for:
    \begin{enumerate}
        \item \textbf{Conventional CDM axions}, detected as a narrow spectral excess in the frequency domain.
        \item \textbf{AQN-induced daily modulation}, identified through power integration over a sidereal day and a phase-coherent temporal analysis.
        \item \textbf{Transient signals from streaming DM}, captured via high-resolution frequency-domain monitoring.
    \end{enumerate}
\end{itemize}

The projected sensitivity of ADAMOS highlights its scientific reach. For CDM axions at \SI{19.95}{\GHz}, the experiment will probe previously unexplored parameter space down to $g_{a\gamma\gamma} = \SI{6.88e-13}{\GeV^{-1}}$ within 30 days of integration. The daily-modulation search will provide a targeted experimental test of the AQN model through its predicted sidereal oscillation and seasonal phase shift. The transient-search mode further extends the discovery potential to gravitational-focusing events, where short-duration density enhancements could make otherwise inaccessible QCD-axion parameter space observable. Ultimately, this continuous long-term operation at a fixed frequency enables unique axion searches that are not accessible to traditional scanning experiments.

Initially, ADAMOS will operate at a fixed resonant frequency to maximize this unique integration time for all three channels. A first aluminum prototype has already been manufactured at the University of Hamburg, and a preliminary characterization is ongoing. In a subsequent phase, future upgrades may implement a tuning capability by introducing a buffer gas. This scheme would allow for narrow mass scans and systematic cross-checks without mechanical tuning elements that could degrade the quality factor or compromise the cavity's thin-shell geometry.

ADAMOS is therefore uniquely positioned to probe a wide range of axion scenarios, enabling it to contribute not only to axion discovery efforts but also to advancing our broader understanding of the dark sector. As the first state-of-the-art cavity haloscope of its kind in Germany, ADAMOS establishes new experimental capabilities in high-frequency axion searches. The experiment's design allows for potential upgrades, including cryogenic operation to reduce system noise and the implementation of quantum-limited amplifiers, which could further enhance sensitivity in subsequent phases of the research program.

\section*{Acknowledgments}

We thank Le Hoang Nguyen for initial discussions and valuable input on the development of this project. This project is funded by the Deutsche Forschungsgemeinschaft (DFG, German Research Foundation) under Germany’s Excellence Strategy – EXC 2121 ``Quantum Universe" – 390833306, and through the DFG funds for major instrumentation grant DFG INST 152/824-1. This article is based upon work from COST Action COSMIC WISPers CA21106, supported by COST (European Cooperation in Science and Technology).

\bibliography{references}

\begin{thebibliography}{52}%
\makeatletter
\providecommand \@ifxundefined [1]{%
 \@ifx{#1\undefined}
}%
\providecommand \@ifnum [1]{%
 \ifnum #1\expandafter \@firstoftwo
 \else \expandafter \@secondoftwo
 \fi
}%
\providecommand \@ifx [1]{%
 \ifx #1\expandafter \@firstoftwo
 \else \expandafter \@secondoftwo
 \fi
}%
\providecommand \natexlab [1]{#1}%
\providecommand \enquote  [1]{``#1''}%
\providecommand \bibnamefont  [1]{#1}%
\providecommand \bibfnamefont [1]{#1}%
\providecommand \citenamefont [1]{#1}%
\providecommand \href@noop [0]{\@secondoftwo}%
\providecommand \href [0]{\begingroup \@sanitize@url \@href}%
\providecommand \@href[1]{\@@startlink{#1}\@@href}%
\providecommand \@@href[1]{\endgroup#1\@@endlink}%
\providecommand \@sanitize@url [0]{\catcode `\\12\catcode `\$12\catcode `\&12\catcode `\#12\catcode `\^12\catcode `\_12\catcode `\%12\relax}%
\providecommand \@@startlink[1]{}%
\providecommand \@@endlink[0]{}%
\providecommand \url  [0]{\begingroup\@sanitize@url \@url }%
\providecommand \@url [1]{\endgroup\@href {#1}{\urlprefix }}%
\providecommand \urlprefix  [0]{URL }%
\providecommand \Eprint [0]{\href }%
\providecommand \doibase [0]{https://doi.org/}%
\providecommand \selectlanguage [0]{\@gobble}%
\providecommand \bibinfo  [0]{\@secondoftwo}%
\providecommand \bibfield  [0]{\@secondoftwo}%
\providecommand \translation [1]{[#1]}%
\providecommand \BibitemOpen [0]{}%
\providecommand \bibitemStop [0]{}%
\providecommand \bibitemNoStop [0]{.\EOS\space}%
\providecommand \EOS [0]{\spacefactor3000\relax}%
\providecommand \BibitemShut  [1]{\csname bibitem#1\endcsname}%
\let\auto@bib@innerbib\@empty
\bibitem [{\citenamefont {Peccei}\ and\ \citenamefont {Quinn}(1977)}]{Peccei_1977_constraints}%
  \BibitemOpen
  \bibfield  {author} {\bibinfo {author} {\bibfnamefont {R.~D.}\ \bibnamefont {Peccei}}\ and\ \bibinfo {author} {\bibfnamefont {H.~R.}\ \bibnamefont {Quinn}},\ }\href {https://doi.org/10.1103/PhysRevD.16.1791} {\bibfield  {journal} {\bibinfo  {journal} {Phys. Rev. D}\ }\textbf {\bibinfo {volume} {16}},\ \bibinfo {pages} {1791} (\bibinfo {year} {1977})}\BibitemShut {NoStop}%
\bibitem [{\citenamefont {Sikivie}(1983)}]{Sikivie_1983_experimental}%
  \BibitemOpen
  \bibfield  {author} {\bibinfo {author} {\bibfnamefont {P.}~\bibnamefont {Sikivie}},\ }\href {https://doi.org/10.1103/PhysRevLett.51.1415} {\bibfield  {journal} {\bibinfo  {journal} {Phys. Rev. Lett.}\ }\textbf {\bibinfo {volume} {51}},\ \bibinfo {pages} {1415} (\bibinfo {year} {1983})}\BibitemShut {NoStop}%
\bibitem [{\citenamefont {Bartram}\ \emph {et~al.}(2021)\citenamefont {Bartram} \emph {et~al.}}]{ADMX_2021_search}%
  \BibitemOpen
  \bibfield  {author} {\bibinfo {author} {\bibfnamefont {C.}~\bibnamefont {Bartram}} \emph {et~al.} (\bibinfo {collaboration} {ADMX}),\ }\href {https://doi.org/10.1103/PhysRevLett.127.261803} {\bibfield  {journal} {\bibinfo  {journal} {Phys. Rev. Lett.}\ }\textbf {\bibinfo {volume} {127}},\ \bibinfo {pages} {261803} (\bibinfo {year} {2021})},\ \Eprint {https://arxiv.org/abs/2110.06096} {arXiv:2110.06096 [hep-ex]} \BibitemShut {NoStop}%
\bibitem [{\citenamefont {Ahn}\ \emph {et~al.}(2024)\citenamefont {Ahn} \emph {et~al.}}]{CAPP_2024_ahn}%
  \BibitemOpen
  \bibfield  {author} {\bibinfo {author} {\bibfnamefont {S.}~\bibnamefont {Ahn}} \emph {et~al.} (\bibinfo {collaboration} {CAPP}),\ }\href {https://doi.org/10.1103/PhysRevX.14.031023} {\bibfield  {journal} {\bibinfo  {journal} {Phys. Rev. X}\ }\textbf {\bibinfo {volume} {14}},\ \bibinfo {pages} {031023} (\bibinfo {year} {2024})},\ \Eprint {https://arxiv.org/abs/2402.12892} {arXiv:2402.12892 [hep-ex]} \BibitemShut {NoStop}%
\bibitem [{\citenamefont {Adair}\ \emph {et~al.}(2022)\citenamefont {Adair}, \citenamefont {Altenm{\"u}ller}, \citenamefont {Anastassopoulos}, \citenamefont {Arguedas~Cuendis}, \citenamefont {Baier}, \citenamefont {Barth}, \citenamefont {Belov}, \citenamefont {Bozicevic}, \citenamefont {Br{\"a}uninger}, \citenamefont {Cantatore}, \citenamefont {Caspers}, \citenamefont {Castel}, \citenamefont {{\c C}etin}, \citenamefont {Chung}, \citenamefont {Choi}, \citenamefont {Choi}, \citenamefont {Dafni}, \citenamefont {Davenport}, \citenamefont {Dermenev}, \citenamefont {Desch}, \citenamefont {D{\"o}brich}, \citenamefont {Fischer}, \citenamefont {Funk}, \citenamefont {Galan}, \citenamefont {Gardikiotis}, \citenamefont {Gninenko}, \citenamefont {Golm}, \citenamefont {Hasinoff}, \citenamefont {Hoffmann}, \citenamefont {D{\'\i}ez~Ib{\'a}{\~n}ez}, \citenamefont {Irastorza}, \citenamefont {Jakov{\v c}i{\'c}}, \citenamefont {Kaminski}, \citenamefont {Karuza}, \citenamefont {Krieger}, \citenamefont {Kutlu}, \citenamefont
  {Laki{\'c}}, \citenamefont {Laurent}, \citenamefont {Lee}, \citenamefont {Lee}, \citenamefont {Luz{\'o}n}, \citenamefont {Malbrunot}, \citenamefont {Margalejo}, \citenamefont {Maroudas}, \citenamefont {Miceli}, \citenamefont {Mirallas}, \citenamefont {Obis}, \citenamefont {{\"O}zbey}, \citenamefont {{\"O}zbozduman}, \citenamefont {Pivovaroff}, \citenamefont {Rosu}, \citenamefont {Ruz}, \citenamefont {Ruiz-Ch{\'o}liz}, \citenamefont {Schmidt}, \citenamefont {Schumann}, \citenamefont {Semertzidis}, \citenamefont {Solanki}, \citenamefont {Stewart}, \citenamefont {Tsagris}, \citenamefont {Vafeiadis}, \citenamefont {Vogel}, \citenamefont {Vretenar}, \citenamefont {Youn},\ and\ \citenamefont {Zioutas}}]{Adair_CAST-CAPP_2022}%
  \BibitemOpen
  \bibfield  {author} {\bibinfo {author} {\bibfnamefont {C.~M.}\ \bibnamefont {Adair}}, \bibinfo {author} {\bibfnamefont {K.}~\bibnamefont {Altenm{\"u}ller}}, \bibinfo {author} {\bibfnamefont {V.}~\bibnamefont {Anastassopoulos}}, \bibinfo {author} {\bibfnamefont {S.}~\bibnamefont {Arguedas~Cuendis}}, \bibinfo {author} {\bibfnamefont {J.}~\bibnamefont {Baier}}, \bibinfo {author} {\bibfnamefont {K.}~\bibnamefont {Barth}}, \bibinfo {author} {\bibfnamefont {A.}~\bibnamefont {Belov}}, \bibinfo {author} {\bibfnamefont {D.}~\bibnamefont {Bozicevic}}, \bibinfo {author} {\bibfnamefont {H.}~\bibnamefont {Br{\"a}uninger}}, \bibinfo {author} {\bibfnamefont {G.}~\bibnamefont {Cantatore}}, \bibinfo {author} {\bibfnamefont {F.}~\bibnamefont {Caspers}}, \bibinfo {author} {\bibfnamefont {J.~F.}\ \bibnamefont {Castel}}, \bibinfo {author} {\bibfnamefont {S.~A.}\ \bibnamefont {{\c C}etin}}, \bibinfo {author} {\bibfnamefont {W.}~\bibnamefont {Chung}}, \bibinfo {author} {\bibfnamefont {H.}~\bibnamefont {Choi}}, \bibinfo {author}
  {\bibfnamefont {J.}~\bibnamefont {Choi}}, \bibinfo {author} {\bibfnamefont {T.}~\bibnamefont {Dafni}}, \bibinfo {author} {\bibfnamefont {M.}~\bibnamefont {Davenport}}, \bibinfo {author} {\bibfnamefont {A.}~\bibnamefont {Dermenev}}, \bibinfo {author} {\bibfnamefont {K.}~\bibnamefont {Desch}}, \bibinfo {author} {\bibfnamefont {B.}~\bibnamefont {D{\"o}brich}}, \bibinfo {author} {\bibfnamefont {H.}~\bibnamefont {Fischer}}, \bibinfo {author} {\bibfnamefont {W.}~\bibnamefont {Funk}}, \bibinfo {author} {\bibfnamefont {J.}~\bibnamefont {Galan}}, \bibinfo {author} {\bibfnamefont {A.}~\bibnamefont {Gardikiotis}}, \bibinfo {author} {\bibfnamefont {S.}~\bibnamefont {Gninenko}}, \bibinfo {author} {\bibfnamefont {J.}~\bibnamefont {Golm}}, \bibinfo {author} {\bibfnamefont {M.~D.}\ \bibnamefont {Hasinoff}}, \bibinfo {author} {\bibfnamefont {D.~H.~H.}\ \bibnamefont {Hoffmann}}, \bibinfo {author} {\bibfnamefont {D.}~\bibnamefont {D{\'\i}ez~Ib{\'a}{\~n}ez}}, \bibinfo {author} {\bibfnamefont {I.~G.}\ \bibnamefont {Irastorza}},
  \bibinfo {author} {\bibfnamefont {K.}~\bibnamefont {Jakov{\v c}i{\'c}}}, \bibinfo {author} {\bibfnamefont {J.}~\bibnamefont {Kaminski}}, \bibinfo {author} {\bibfnamefont {M.}~\bibnamefont {Karuza}}, \bibinfo {author} {\bibfnamefont {C.}~\bibnamefont {Krieger}}, \bibinfo {author} {\bibfnamefont {{\c C}.}~\bibnamefont {Kutlu}}, \bibinfo {author} {\bibfnamefont {B.}~\bibnamefont {Laki{\'c}}}, \bibinfo {author} {\bibfnamefont {J.~M.}\ \bibnamefont {Laurent}}, \bibinfo {author} {\bibfnamefont {J.}~\bibnamefont {Lee}}, \bibinfo {author} {\bibfnamefont {S.}~\bibnamefont {Lee}}, \bibinfo {author} {\bibfnamefont {G.}~\bibnamefont {Luz{\'o}n}}, \bibinfo {author} {\bibfnamefont {C.}~\bibnamefont {Malbrunot}}, \bibinfo {author} {\bibfnamefont {C.}~\bibnamefont {Margalejo}}, \bibinfo {author} {\bibfnamefont {M.}~\bibnamefont {Maroudas}}, \bibinfo {author} {\bibfnamefont {L.}~\bibnamefont {Miceli}}, \bibinfo {author} {\bibfnamefont {H.}~\bibnamefont {Mirallas}}, \bibinfo {author} {\bibfnamefont {L.}~\bibnamefont {Obis}},
  \bibinfo {author} {\bibfnamefont {A.}~\bibnamefont {{\"O}zbey}}, \bibinfo {author} {\bibfnamefont {K.}~\bibnamefont {{\"O}zbozduman}}, \bibinfo {author} {\bibfnamefont {M.~J.}\ \bibnamefont {Pivovaroff}}, \bibinfo {author} {\bibfnamefont {M.}~\bibnamefont {Rosu}}, \bibinfo {author} {\bibfnamefont {J.}~\bibnamefont {Ruz}}, \bibinfo {author} {\bibfnamefont {E.}~\bibnamefont {Ruiz-Ch{\'o}liz}}, \bibinfo {author} {\bibfnamefont {S.}~\bibnamefont {Schmidt}}, \bibinfo {author} {\bibfnamefont {M.}~\bibnamefont {Schumann}}, \bibinfo {author} {\bibfnamefont {Y.~K.}\ \bibnamefont {Semertzidis}}, \bibinfo {author} {\bibfnamefont {S.~K.}\ \bibnamefont {Solanki}}, \bibinfo {author} {\bibfnamefont {L.}~\bibnamefont {Stewart}}, \bibinfo {author} {\bibfnamefont {I.}~\bibnamefont {Tsagris}}, \bibinfo {author} {\bibfnamefont {T.}~\bibnamefont {Vafeiadis}}, \bibinfo {author} {\bibfnamefont {J.~K.}\ \bibnamefont {Vogel}}, \bibinfo {author} {\bibfnamefont {M.}~\bibnamefont {Vretenar}}, \bibinfo {author} {\bibfnamefont
  {S.}~\bibnamefont {Youn}},\ and\ \bibinfo {author} {\bibfnamefont {K.}~\bibnamefont {Zioutas}},\ }\href {https://doi.org/10.1038/s41467-022-33913-6} {\bibfield  {journal} {\bibinfo  {journal} {Nature Communications}\ }\textbf {\bibinfo {volume} {13}},\ \bibinfo {pages} {6180} (\bibinfo {year} {2022})}\BibitemShut {NoStop}%
\bibitem [{\citenamefont {Borsanyi}\ \emph {et~al.}(2016)\citenamefont {Borsanyi}, \citenamefont {Fodor}, \citenamefont {Guenther}, \citenamefont {Kampert}, \citenamefont {Katz}, \citenamefont {Kawanai}, \citenamefont {Kovacs}, \citenamefont {Mages}, \citenamefont {Pasztor}, \citenamefont {Pittler}, \citenamefont {Redondo}, \citenamefont {Ringwald},\ and\ \citenamefont {Szabo}}]{Borsanyi_2016_mass}%
  \BibitemOpen
  \bibfield  {author} {\bibinfo {author} {\bibfnamefont {S.}~\bibnamefont {Borsanyi}}, \bibinfo {author} {\bibfnamefont {Z.}~\bibnamefont {Fodor}}, \bibinfo {author} {\bibfnamefont {J.}~\bibnamefont {Guenther}}, \bibinfo {author} {\bibfnamefont {K.~H.}\ \bibnamefont {Kampert}}, \bibinfo {author} {\bibfnamefont {S.~D.}\ \bibnamefont {Katz}}, \bibinfo {author} {\bibfnamefont {T.}~\bibnamefont {Kawanai}}, \bibinfo {author} {\bibfnamefont {T.~G.}\ \bibnamefont {Kovacs}}, \bibinfo {author} {\bibfnamefont {S.~W.}\ \bibnamefont {Mages}}, \bibinfo {author} {\bibfnamefont {A.}~\bibnamefont {Pasztor}}, \bibinfo {author} {\bibfnamefont {F.}~\bibnamefont {Pittler}}, \bibinfo {author} {\bibfnamefont {J.}~\bibnamefont {Redondo}}, \bibinfo {author} {\bibfnamefont {A.}~\bibnamefont {Ringwald}},\ and\ \bibinfo {author} {\bibfnamefont {K.~K.}\ \bibnamefont {Szabo}},\ }\href {https://doi.org/10.1038/nature20115} {\bibfield  {journal} {\bibinfo  {journal} {Nature}\ }\textbf {\bibinfo {volume} {539}},\ \bibinfo {pages} {69}
  (\bibinfo {year} {2016})}\BibitemShut {NoStop}%
\bibitem [{\citenamefont {Zhitnitsky}(2003)}]{Zhitnitsky_2003_aqn}%
  \BibitemOpen
  \bibfield  {author} {\bibinfo {author} {\bibfnamefont {A.~R.}\ \bibnamefont {Zhitnitsky}},\ }\href {https://doi.org/10.1088/1475-7516/2003/10/010} {\bibfield  {journal} {\bibinfo  {journal} {Journal of Cosmology and Astroparticle Physics}\ }\textbf {\bibinfo {volume} {2003}}\bibinfo  {number} { (10)},\ \bibinfo {pages} {010}}\BibitemShut {NoStop}%
\bibitem [{\citenamefont {Liang}\ \emph {et~al.}(2020)\citenamefont {Liang}, \citenamefont {Mead}, \citenamefont {Siddiqui}, \citenamefont {Waerbeke},\ and\ \citenamefont {Zhitnitsky}}]{Liang_aqn_2020}%
  \BibitemOpen
\bibfield  {number} {  }\bibfield  {author} {\bibinfo {author} {\bibfnamefont {X.}~\bibnamefont {Liang}}, \bibinfo {author} {\bibfnamefont {A.}~\bibnamefont {Mead}}, \bibinfo {author} {\bibfnamefont {M.~S.~R.}\ \bibnamefont {Siddiqui}}, \bibinfo {author} {\bibfnamefont {L.~V.}\ \bibnamefont {Waerbeke}},\ and\ \bibinfo {author} {\bibfnamefont {A.}~\bibnamefont {Zhitnitsky}},\ }\bibfield  {journal} {\bibinfo  {journal} {Physical Review D}\ }\textbf {\bibinfo {volume} {101}},\ \href {https://doi.org/10.1103/PhysRevD.101.043512} {10.1103/PhysRevD.101.043512} (\bibinfo {year} {2020}),\ \bibinfo {note} {arXiv:1908.04675 [astro-ph]}\BibitemShut {NoStop}%
\bibitem [{\citenamefont {Budker}\ \emph {et~al.}(2020)\citenamefont {Budker}, \citenamefont {Flambaum}, \citenamefont {Liang},\ and\ \citenamefont {Zhitnitsky}}]{Budker_AQN_2020}%
  \BibitemOpen
  \bibfield  {author} {\bibinfo {author} {\bibfnamefont {D.}~\bibnamefont {Budker}}, \bibinfo {author} {\bibfnamefont {V.~V.}\ \bibnamefont {Flambaum}}, \bibinfo {author} {\bibfnamefont {X.}~\bibnamefont {Liang}},\ and\ \bibinfo {author} {\bibfnamefont {A.}~\bibnamefont {Zhitnitsky}},\ }\href {https://doi.org/10.1103/PhysRevD.101.043012} {\bibfield  {journal} {\bibinfo  {journal} {Phys. Rev. D}\ }\textbf {\bibinfo {volume} {101}},\ \bibinfo {pages} {043012} (\bibinfo {year} {2020})}\BibitemShut {NoStop}%
\bibitem [{\citenamefont {Caspers}\ \emph {et~al.}(2025)\citenamefont {Caspers}, \citenamefont {Adair}, \citenamefont {Altenm\"uller}, \citenamefont {Anastassopoulos}, \citenamefont {Arguedas~Cuendis}, \citenamefont {Baier}, \citenamefont {Barth}, \citenamefont {Belov}, \citenamefont {Bozicevic}, \citenamefont {Br\"auninger}, \citenamefont {Cantatore}, \citenamefont {Castel}, \citenamefont {\ifmmode~\mbox{\c{C}}\else \c{C}\fi{}etin}, \citenamefont {Chung}, \citenamefont {Choi}, \citenamefont {Choi}, \citenamefont {Dafni}, \citenamefont {Davenport}, \citenamefont {Dermenev}, \citenamefont {Desch}, \citenamefont {D\"obrich}, \citenamefont {Fischer}, \citenamefont {Funk}, \citenamefont {Galan}, \citenamefont {Gardikiotis}, \citenamefont {Gninenko}, \citenamefont {Golm}, \citenamefont {Hasinoff}, \citenamefont {Hoffmann}, \citenamefont {D\'{\i}ez Ib\'a\~nez}, \citenamefont {Irastorza}, \citenamefont {Jakov\ifmmode \check{c}\else \v{c}\fi{}i\ifmmode~\acute{c}\else \'{c}\fi{}}, \citenamefont {Kaminski},
  \citenamefont {Karuza}, \citenamefont {Krieger}, \citenamefont {Kutlu}, \citenamefont {Laki\ifmmode~\acute{c}\else \'{c}\fi{}}, \citenamefont {Laurent}, \citenamefont {Lee}, \citenamefont {Lee}, \citenamefont {Luz\'on}, \citenamefont {Margalejo}, \citenamefont {Maroudas}, \citenamefont {Miceli}, \citenamefont {Mirallas}, \citenamefont {Obis}, \citenamefont {\"Ozbey}, \citenamefont {\"Ozbozduman}, \citenamefont {Pivovaroff}, \citenamefont {Rosu}, \citenamefont {Ruz}, \citenamefont {Ruiz-Ch\'oliz}, \citenamefont {Schmidt}, \citenamefont {Semertzidis}, \citenamefont {Solanki}, \citenamefont {Stewart}, \citenamefont {Tsagris}, \citenamefont {Vafeiadis}, \citenamefont {Vogel}, \citenamefont {Vretenar}, \citenamefont {Youn}, \citenamefont {Zhitnitsky},\ and\ \citenamefont {Zioutas}}]{Caspers_daily_2025}%
  \BibitemOpen
  \bibfield  {author} {\bibinfo {author} {\bibfnamefont {F.}~\bibnamefont {Caspers}}, \bibinfo {author} {\bibfnamefont {C.~M.}\ \bibnamefont {Adair}}, \bibinfo {author} {\bibfnamefont {K.}~\bibnamefont {Altenm\"uller}}, \bibinfo {author} {\bibfnamefont {V.}~\bibnamefont {Anastassopoulos}}, \bibinfo {author} {\bibfnamefont {S.}~\bibnamefont {Arguedas~Cuendis}}, \bibinfo {author} {\bibfnamefont {J.}~\bibnamefont {Baier}}, \bibinfo {author} {\bibfnamefont {K.}~\bibnamefont {Barth}}, \bibinfo {author} {\bibfnamefont {A.}~\bibnamefont {Belov}}, \bibinfo {author} {\bibfnamefont {D.}~\bibnamefont {Bozicevic}}, \bibinfo {author} {\bibfnamefont {H.}~\bibnamefont {Br\"auninger}}, \bibinfo {author} {\bibfnamefont {G.}~\bibnamefont {Cantatore}}, \bibinfo {author} {\bibfnamefont {J.~F.}\ \bibnamefont {Castel}}, \bibinfo {author} {\bibfnamefont {S.~A.}\ \bibnamefont {\ifmmode~\mbox{\c{C}}\else \c{C}\fi{}etin}}, \bibinfo {author} {\bibfnamefont {W.}~\bibnamefont {Chung}}, \bibinfo {author} {\bibfnamefont {H.}~\bibnamefont
  {Choi}}, \bibinfo {author} {\bibfnamefont {J.}~\bibnamefont {Choi}}, \bibinfo {author} {\bibfnamefont {T.}~\bibnamefont {Dafni}}, \bibinfo {author} {\bibfnamefont {M.}~\bibnamefont {Davenport}}, \bibinfo {author} {\bibfnamefont {A.}~\bibnamefont {Dermenev}}, \bibinfo {author} {\bibfnamefont {K.}~\bibnamefont {Desch}}, \bibinfo {author} {\bibfnamefont {B.}~\bibnamefont {D\"obrich}}, \bibinfo {author} {\bibfnamefont {H.}~\bibnamefont {Fischer}}, \bibinfo {author} {\bibfnamefont {W.}~\bibnamefont {Funk}}, \bibinfo {author} {\bibfnamefont {J.}~\bibnamefont {Galan}}, \bibinfo {author} {\bibfnamefont {A.}~\bibnamefont {Gardikiotis}}, \bibinfo {author} {\bibfnamefont {S.}~\bibnamefont {Gninenko}}, \bibinfo {author} {\bibfnamefont {J.}~\bibnamefont {Golm}}, \bibinfo {author} {\bibfnamefont {M.~D.}\ \bibnamefont {Hasinoff}}, \bibinfo {author} {\bibfnamefont {D.~H.~H.}\ \bibnamefont {Hoffmann}}, \bibinfo {author} {\bibfnamefont {D.}~\bibnamefont {D\'{\i}ez Ib\'a\~nez}}, \bibinfo {author} {\bibfnamefont {I.~G.}\
  \bibnamefont {Irastorza}}, \bibinfo {author} {\bibfnamefont {K.}~\bibnamefont {Jakov\ifmmode \check{c}\else \v{c}\fi{}i\ifmmode~\acute{c}\else \'{c}\fi{}}}, \bibinfo {author} {\bibfnamefont {J.}~\bibnamefont {Kaminski}}, \bibinfo {author} {\bibfnamefont {M.}~\bibnamefont {Karuza}}, \bibinfo {author} {\bibfnamefont {C.}~\bibnamefont {Krieger}}, \bibinfo {author} {\bibfnamefont {i.~m.~c.}\ \bibnamefont {Kutlu}}, \bibinfo {author} {\bibfnamefont {B.}~\bibnamefont {Laki\ifmmode~\acute{c}\else \'{c}\fi{}}}, \bibinfo {author} {\bibfnamefont {J.~M.}\ \bibnamefont {Laurent}}, \bibinfo {author} {\bibfnamefont {J.}~\bibnamefont {Lee}}, \bibinfo {author} {\bibfnamefont {S.}~\bibnamefont {Lee}}, \bibinfo {author} {\bibfnamefont {G.}~\bibnamefont {Luz\'on}}, \bibinfo {author} {\bibfnamefont {C.}~\bibnamefont {Margalejo}}, \bibinfo {author} {\bibfnamefont {M.}~\bibnamefont {Maroudas}}, \bibinfo {author} {\bibfnamefont {L.}~\bibnamefont {Miceli}}, \bibinfo {author} {\bibfnamefont {H.}~\bibnamefont {Mirallas}}, \bibinfo
  {author} {\bibfnamefont {L.}~\bibnamefont {Obis}}, \bibinfo {author} {\bibfnamefont {A.}~\bibnamefont {\"Ozbey}}, \bibinfo {author} {\bibfnamefont {K.}~\bibnamefont {\"Ozbozduman}}, \bibinfo {author} {\bibfnamefont {M.~J.}\ \bibnamefont {Pivovaroff}}, \bibinfo {author} {\bibfnamefont {M.}~\bibnamefont {Rosu}}, \bibinfo {author} {\bibfnamefont {J.}~\bibnamefont {Ruz}}, \bibinfo {author} {\bibfnamefont {E.}~\bibnamefont {Ruiz-Ch\'oliz}}, \bibinfo {author} {\bibfnamefont {S.}~\bibnamefont {Schmidt}}, \bibinfo {author} {\bibfnamefont {Y.~K.}\ \bibnamefont {Semertzidis}}, \bibinfo {author} {\bibfnamefont {S.~K.}\ \bibnamefont {Solanki}}, \bibinfo {author} {\bibfnamefont {L.}~\bibnamefont {Stewart}}, \bibinfo {author} {\bibfnamefont {I.}~\bibnamefont {Tsagris}}, \bibinfo {author} {\bibfnamefont {T.}~\bibnamefont {Vafeiadis}}, \bibinfo {author} {\bibfnamefont {J.~K.}\ \bibnamefont {Vogel}}, \bibinfo {author} {\bibfnamefont {M.}~\bibnamefont {Vretenar}}, \bibinfo {author} {\bibfnamefont {S.}~\bibnamefont {Youn}},
  \bibinfo {author} {\bibfnamefont {A.}~\bibnamefont {Zhitnitsky}},\ and\ \bibinfo {author} {\bibfnamefont {K.}~\bibnamefont {Zioutas}},\ }\href {https://doi.org/10.1103/PhysRevD.111.082009} {\bibfield  {journal} {\bibinfo  {journal} {Phys. Rev. D}\ }\textbf {\bibinfo {volume} {111}},\ \bibinfo {pages} {082009} (\bibinfo {year} {2025})}\BibitemShut {NoStop}%
\bibitem [{\citenamefont {Kim}\ \emph {et~al.}(2025)\citenamefont {Kim}, \citenamefont {Ahn}, \citenamefont {Lee}, \citenamefont {Ahn}, \citenamefont {Chung}, \citenamefont {Kwon}, \citenamefont {Matlashov}, \citenamefont {Uchaikin}, \citenamefont {Ivanov}, \citenamefont {Lee}, \citenamefont {van Loo}, \citenamefont {Nakamura}, \citenamefont {Park}, \citenamefont {Byun}, \citenamefont {Oh}, \citenamefont {Youm},\ and\ \citenamefont {Semertzidis}}]{CAPP_2025_first_aqn}%
  \BibitemOpen
  \bibfield  {author} {\bibinfo {author} {\bibfnamefont {J.}~\bibnamefont {Kim}}, \bibinfo {author} {\bibfnamefont {D.}~\bibnamefont {Ahn}}, \bibinfo {author} {\bibfnamefont {S.}~\bibnamefont {Lee}}, \bibinfo {author} {\bibfnamefont {S.}~\bibnamefont {Ahn}}, \bibinfo {author} {\bibfnamefont {W.}~\bibnamefont {Chung}}, \bibinfo {author} {\bibfnamefont {O.}~\bibnamefont {Kwon}}, \bibinfo {author} {\bibfnamefont {A.}~\bibnamefont {Matlashov}}, \bibinfo {author} {\bibfnamefont {S.~V.}\ \bibnamefont {Uchaikin}}, \bibinfo {author} {\bibfnamefont {B.~I.}\ \bibnamefont {Ivanov}}, \bibinfo {author} {\bibfnamefont {K.~W.}\ \bibnamefont {Lee}}, \bibinfo {author} {\bibfnamefont {A.~F.}\ \bibnamefont {van Loo}}, \bibinfo {author} {\bibfnamefont {Y.}~\bibnamefont {Nakamura}}, \bibinfo {author} {\bibfnamefont {S.}~\bibnamefont {Park}}, \bibinfo {author} {\bibfnamefont {H.}~\bibnamefont {Byun}}, \bibinfo {author} {\bibfnamefont {S.}~\bibnamefont {Oh}}, \bibinfo {author} {\bibfnamefont {D.}~\bibnamefont {Youm}},\ and\
  \bibinfo {author} {\bibfnamefont {Y.~K.}\ \bibnamefont {Semertzidis}},\ }\href {https://doi.org/10.1103/742y-g3fd} {\bibfield  {journal} {\bibinfo  {journal} {Phys. Rev. D}\ }\textbf {\bibinfo {volume} {112}},\ \bibinfo {pages} {L121305} (\bibinfo {year} {2025})}\BibitemShut {NoStop}%
\bibitem [{\citenamefont {Nitta}\ \emph {et~al.}(2023)\citenamefont {Nitta} \emph {et~al.}}]{ADMX_2023_background}%
  \BibitemOpen
  \bibfield  {author} {\bibinfo {author} {\bibfnamefont {T.}~\bibnamefont {Nitta}} \emph {et~al.} (\bibinfo {collaboration} {ADMX}),\ }\href {https://doi.org/10.1103/PhysRevLett.131.101002} {\bibfield  {journal} {\bibinfo  {journal} {Phys. Rev. Lett.}\ }\textbf {\bibinfo {volume} {131}},\ \bibinfo {pages} {101002} (\bibinfo {year} {2023})},\ \Eprint {https://arxiv.org/abs/2303.06282} {arXiv:2303.06282 [hep-ex]} \BibitemShut {NoStop}%
\bibitem [{\citenamefont {Vogelsberger}\ and\ \citenamefont {White}(2011)}]{Vogelsberger_2011_streams}%
  \BibitemOpen
  \bibfield  {author} {\bibinfo {author} {\bibfnamefont {M.}~\bibnamefont {Vogelsberger}}\ and\ \bibinfo {author} {\bibfnamefont {S.~D.~M.}\ \bibnamefont {White}},\ }\href {https://doi.org/10.1111/j.1365-2966.2011.18224.x} {\bibfield  {journal} {\bibinfo  {journal} {Monthly Notices of the Royal Astronomical Society}\ }\textbf {\bibinfo {volume} {413}},\ \bibinfo {pages} {1419} (\bibinfo {year} {2011})},\ \Eprint {https://arxiv.org/abs/https://academic.oup.com/mnras/article-pdf/413/2/1419/18595473/mnras0413-1419.pdf} {https://academic.oup.com/mnras/article-pdf/413/2/1419/18595473/mnras0413-1419.pdf} \BibitemShut {NoStop}%
\bibitem [{\citenamefont {{Tkachev}}(1991)}]{Tkachev_1991_minicluster}%
  \BibitemOpen
  \bibfield  {author} {\bibinfo {author} {\bibfnamefont {I.~I.}\ \bibnamefont {{Tkachev}}},\ }\href {https://doi.org/10.1016/0370-2693(91)90330-S} {\bibfield  {journal} {\bibinfo  {journal} {Physics Letters B}\ }\textbf {\bibinfo {volume} {261}},\ \bibinfo {pages} {289} (\bibinfo {year} {1991})}\BibitemShut {NoStop}%
\bibitem [{\citenamefont {Kryemadhi}\ \emph {et~al.}(2023)\citenamefont {Kryemadhi}, \citenamefont {Maroudas}, \citenamefont {Mastronikolis},\ and\ \citenamefont {Zioutas}}]{Kryemadhi_gravitational_2023}%
  \BibitemOpen
  \bibfield  {author} {\bibinfo {author} {\bibfnamefont {A.}~\bibnamefont {Kryemadhi}}, \bibinfo {author} {\bibfnamefont {M.}~\bibnamefont {Maroudas}}, \bibinfo {author} {\bibfnamefont {A.}~\bibnamefont {Mastronikolis}},\ and\ \bibinfo {author} {\bibfnamefont {K.}~\bibnamefont {Zioutas}},\ }\href {https://doi.org/10.1103/PhysRevD.108.123043} {\bibfield  {journal} {\bibinfo  {journal} {Phys. Rev. D}\ }\textbf {\bibinfo {volume} {108}},\ \bibinfo {pages} {123043} (\bibinfo {year} {2023})}\BibitemShut {NoStop}%
\bibitem [{\citenamefont {O'Hare}\ \emph {et~al.}(2024)\citenamefont {O'Hare}, \citenamefont {Pierobon},\ and\ \citenamefont {Redondo}}]{OHare_2023_Minicluster}%
  \BibitemOpen
  \bibfield  {author} {\bibinfo {author} {\bibfnamefont {C.~A.~J.}\ \bibnamefont {O'Hare}}, \bibinfo {author} {\bibfnamefont {G.}~\bibnamefont {Pierobon}},\ and\ \bibinfo {author} {\bibfnamefont {J.}~\bibnamefont {Redondo}},\ }\href {https://doi.org/10.1103/PhysRevLett.133.081001} {\bibfield  {journal} {\bibinfo  {journal} {Phys. Rev. Lett.}\ }\textbf {\bibinfo {volume} {133}},\ \bibinfo {pages} {081001} (\bibinfo {year} {2024})},\ \Eprint {https://arxiv.org/abs/2311.17367} {arXiv:2311.17367 [hep-ph]} \BibitemShut {NoStop}%
\bibitem [{\citenamefont {Kuo}(2020)}]{Kuo_2019_large}%
  \BibitemOpen
  \bibfield  {author} {\bibinfo {author} {\bibfnamefont {C.-L.}\ \bibnamefont {Kuo}},\ }\href {https://doi.org/10.1088/1475-7516/2020/06/010} {\bibfield  {journal} {\bibinfo  {journal} {JCAP}\ }\textbf {\bibinfo {volume} {06}},\ \bibinfo {pages} {010}},\ \Eprint {https://arxiv.org/abs/1910.04156} {arXiv:1910.04156 [physics.ins-det]} \BibitemShut {NoStop}%
\bibitem [{\citenamefont {Turner}(1990)}]{turner_1990_periodic}%
  \BibitemOpen
  \bibfield  {author} {\bibinfo {author} {\bibfnamefont {M.~S.}\ \bibnamefont {Turner}},\ }\href {https://doi.org/10.1103/PhysRevD.42.3572} {\bibfield  {journal} {\bibinfo  {journal} {Phys. Rev. D}\ }\textbf {\bibinfo {volume} {42}},\ \bibinfo {pages} {3572} (\bibinfo {year} {1990})}\BibitemShut {NoStop}%
\bibitem [{\citenamefont {Brubaker}\ \emph {et~al.}(2017)\citenamefont {Brubaker}, \citenamefont {Zhong}, \citenamefont {Lamoreaux}, \citenamefont {Lehnert},\ and\ \citenamefont {van Bibber}}]{haystac_2017_analysis}%
  \BibitemOpen
  \bibfield  {author} {\bibinfo {author} {\bibfnamefont {B.~M.}\ \bibnamefont {Brubaker}}, \bibinfo {author} {\bibfnamefont {L.}~\bibnamefont {Zhong}}, \bibinfo {author} {\bibfnamefont {S.~K.}\ \bibnamefont {Lamoreaux}}, \bibinfo {author} {\bibfnamefont {K.~W.}\ \bibnamefont {Lehnert}},\ and\ \bibinfo {author} {\bibfnamefont {K.~A.}\ \bibnamefont {van Bibber}},\ }\href {https://doi.org/10.1103/PhysRevD.96.123008} {\bibfield  {journal} {\bibinfo  {journal} {Phys. Rev. D}\ }\textbf {\bibinfo {volume} {96}},\ \bibinfo {pages} {123008} (\bibinfo {year} {2017})}\BibitemShut {NoStop}%
\bibitem [{\citenamefont {Ahn}\ \emph {et~al.}(2022)\citenamefont {Ahn}, \citenamefont {Kwon}, \citenamefont {Chung}, \citenamefont {Jang}, \citenamefont {Lee}, \citenamefont {Lee}, \citenamefont {Youn}, \citenamefont {Byun}, \citenamefont {Youm},\ and\ \citenamefont {Semertzidis}}]{capp_2022_hts}%
  \BibitemOpen
  \bibfield  {author} {\bibinfo {author} {\bibfnamefont {D.}~\bibnamefont {Ahn}}, \bibinfo {author} {\bibfnamefont {O.}~\bibnamefont {Kwon}}, \bibinfo {author} {\bibfnamefont {W.}~\bibnamefont {Chung}}, \bibinfo {author} {\bibfnamefont {W.}~\bibnamefont {Jang}}, \bibinfo {author} {\bibfnamefont {D.}~\bibnamefont {Lee}}, \bibinfo {author} {\bibfnamefont {J.}~\bibnamefont {Lee}}, \bibinfo {author} {\bibfnamefont {S.~W.}\ \bibnamefont {Youn}}, \bibinfo {author} {\bibfnamefont {H.}~\bibnamefont {Byun}}, \bibinfo {author} {\bibfnamefont {D.}~\bibnamefont {Youm}},\ and\ \bibinfo {author} {\bibfnamefont {Y.~K.}\ \bibnamefont {Semertzidis}},\ }\href {https://doi.org/10.1103/PhysRevApplied.17.L061005} {\bibfield  {journal} {\bibinfo  {journal} {Phys. Rev. Appl.}\ }\textbf {\bibinfo {volume} {17}},\ \bibinfo {pages} {L061005} (\bibinfo {year} {2022})}\BibitemShut {NoStop}%
\bibitem [{\citenamefont {Goryachev}\ \emph {et~al.}(2018)\citenamefont {Goryachev}, \citenamefont {Mcallister},\ and\ \citenamefont {Tobar}}]{Goryachev_2017_arrays}%
  \BibitemOpen
  \bibfield  {author} {\bibinfo {author} {\bibfnamefont {M.}~\bibnamefont {Goryachev}}, \bibinfo {author} {\bibfnamefont {B.~T.}\ \bibnamefont {Mcallister}},\ and\ \bibinfo {author} {\bibfnamefont {M.~E.}\ \bibnamefont {Tobar}},\ }\href {https://doi.org/10.1016/j.physleta.2017.09.016} {\bibfield  {journal} {\bibinfo  {journal} {Phys. Lett. A}\ }\textbf {\bibinfo {volume} {382}},\ \bibinfo {pages} {2199} (\bibinfo {year} {2018})},\ \Eprint {https://arxiv.org/abs/1703.07207} {arXiv:1703.07207 [physics.ins-det]} \BibitemShut {NoStop}%
\bibitem [{\citenamefont {Melc{\'o}n}\ \emph {et~al.}(2020)\citenamefont {Melc{\'o}n}, \citenamefont {Cuendis}, \citenamefont {Cogollos}, \citenamefont {D{\'\i}az-Morcillo}, \citenamefont {D{\"o}brich}, \citenamefont {Gallego}, \citenamefont {Barcel{\'o}}, \citenamefont {Gimeno}, \citenamefont {Golm}, \citenamefont {Irastorza}, \citenamefont {Lozano-Guerrero}, \citenamefont {Malbrunot}, \citenamefont {Millar}, \citenamefont {Navarro}, \citenamefont {Garay}, \citenamefont {Redondo},\ and\ \citenamefont {Wuensch}}]{rades_2020_scalable}%
  \BibitemOpen
  \bibfield  {author} {\bibinfo {author} {\bibfnamefont {A.~{\'A}.}\ \bibnamefont {Melc{\'o}n}}, \bibinfo {author} {\bibfnamefont {S.~A.}\ \bibnamefont {Cuendis}}, \bibinfo {author} {\bibfnamefont {C.}~\bibnamefont {Cogollos}}, \bibinfo {author} {\bibfnamefont {A.}~\bibnamefont {D{\'\i}az-Morcillo}}, \bibinfo {author} {\bibfnamefont {B.}~\bibnamefont {D{\"o}brich}}, \bibinfo {author} {\bibfnamefont {J.~D.}\ \bibnamefont {Gallego}}, \bibinfo {author} {\bibfnamefont {J.~M.~G.}\ \bibnamefont {Barcel{\'o}}}, \bibinfo {author} {\bibfnamefont {B.}~\bibnamefont {Gimeno}}, \bibinfo {author} {\bibfnamefont {J.}~\bibnamefont {Golm}}, \bibinfo {author} {\bibfnamefont {I.~G.}\ \bibnamefont {Irastorza}}, \bibinfo {author} {\bibfnamefont {A.~J.}\ \bibnamefont {Lozano-Guerrero}}, \bibinfo {author} {\bibfnamefont {C.}~\bibnamefont {Malbrunot}}, \bibinfo {author} {\bibfnamefont {A.}~\bibnamefont {Millar}}, \bibinfo {author} {\bibfnamefont {P.}~\bibnamefont {Navarro}}, \bibinfo {author} {\bibfnamefont {C.~P.}\ \bibnamefont
  {Garay}}, \bibinfo {author} {\bibfnamefont {J.}~\bibnamefont {Redondo}},\ and\ \bibinfo {author} {\bibfnamefont {W.}~\bibnamefont {Wuensch}},\ }\href {https://doi.org/10.1007/JHEP07(2020)084} {\bibfield  {journal} {\bibinfo  {journal} {Journal of High Energy Physics}\ }\textbf {\bibinfo {volume} {2020}},\ \bibinfo {pages} {84} (\bibinfo {year} {2020})}\BibitemShut {NoStop}%
\bibitem [{\citenamefont {Jeong}\ \emph {et~al.}(2023)\citenamefont {Jeong}, \citenamefont {Youn},\ and\ \citenamefont {Kim}}]{Junu_2023_multicell}%
  \BibitemOpen
  \bibfield  {author} {\bibinfo {author} {\bibfnamefont {J.}~\bibnamefont {Jeong}}, \bibinfo {author} {\bibfnamefont {S.}~\bibnamefont {Youn}},\ and\ \bibinfo {author} {\bibfnamefont {J.~E.}\ \bibnamefont {Kim}},\ }\href {https://doi.org/https://doi.org/10.1016/j.nima.2023.168327} {\bibfield  {journal} {\bibinfo  {journal} {Nuclear Instruments and Methods in Physics Research Section A: Accelerators, Spectrometers, Detectors and Associated Equipment}\ }\textbf {\bibinfo {volume} {1053}},\ \bibinfo {pages} {168327} (\bibinfo {year} {2023})}\BibitemShut {NoStop}%
\bibitem [{\citenamefont {Horns}\ \emph {et~al.}(2013)\citenamefont {Horns}, \citenamefont {Jaeckel}, \citenamefont {Lindner}, \citenamefont {Lobanov}, \citenamefont {Redondo},\ and\ \citenamefont {Ringwald}}]{horns_2013_dish}%
  \BibitemOpen
  \bibfield  {author} {\bibinfo {author} {\bibfnamefont {D.}~\bibnamefont {Horns}}, \bibinfo {author} {\bibfnamefont {J.}~\bibnamefont {Jaeckel}}, \bibinfo {author} {\bibfnamefont {A.}~\bibnamefont {Lindner}}, \bibinfo {author} {\bibfnamefont {A.}~\bibnamefont {Lobanov}}, \bibinfo {author} {\bibfnamefont {J.}~\bibnamefont {Redondo}},\ and\ \bibinfo {author} {\bibfnamefont {A.}~\bibnamefont {Ringwald}},\ }\href {https://doi.org/10.1088/1475-7516/2013/04/016} {\bibfield  {journal} {\bibinfo  {journal} {Journal of Cosmology and Astroparticle Physics}\ }\textbf {\bibinfo {volume} {2013}}\bibinfo  {number} { (04)},\ \bibinfo {pages} {016}}\BibitemShut {NoStop}%
\bibitem [{\citenamefont {Caldwell}\ \emph {et~al.}(2017)\citenamefont {Caldwell}, \citenamefont {Dvali}, \citenamefont {Majorovits}, \citenamefont {Millar}, \citenamefont {Raffelt}, \citenamefont {Redondo}, \citenamefont {Reimann}, \citenamefont {Simon},\ and\ \citenamefont {Steffen}}]{madmax_2017_dielectric}%
  \BibitemOpen
\bibfield  {number} {  }\bibfield  {author} {\bibinfo {author} {\bibfnamefont {A.}~\bibnamefont {Caldwell}}, \bibinfo {author} {\bibfnamefont {G.}~\bibnamefont {Dvali}}, \bibinfo {author} {\bibfnamefont {B.}~\bibnamefont {Majorovits}}, \bibinfo {author} {\bibfnamefont {A.}~\bibnamefont {Millar}}, \bibinfo {author} {\bibfnamefont {G.}~\bibnamefont {Raffelt}}, \bibinfo {author} {\bibfnamefont {J.}~\bibnamefont {Redondo}}, \bibinfo {author} {\bibfnamefont {O.}~\bibnamefont {Reimann}}, \bibinfo {author} {\bibfnamefont {F.}~\bibnamefont {Simon}},\ and\ \bibinfo {author} {\bibfnamefont {F.}~\bibnamefont {Steffen}} (\bibinfo {collaboration} {MADMAX Working Group}),\ }\href {https://doi.org/10.1103/PhysRevLett.118.091801} {\bibfield  {journal} {\bibinfo  {journal} {Phys. Rev. Lett.}\ }\textbf {\bibinfo {volume} {118}},\ \bibinfo {pages} {091801} (\bibinfo {year} {2017})}\BibitemShut {NoStop}%
\bibitem [{\citenamefont {Millar}\ \emph {et~al.}(2023)\citenamefont {Millar}, \citenamefont {Anlage}, \citenamefont {Balafendiev}, \citenamefont {Belov}, \citenamefont {van Bibber}, \citenamefont {Conrad}, \citenamefont {Demarteau}, \citenamefont {Droster}, \citenamefont {Dunne}, \citenamefont {Rosso}, \citenamefont {Gudmundsson}, \citenamefont {Jackson}, \citenamefont {Kaur}, \citenamefont {Klaesson}, \citenamefont {Kowitt}, \citenamefont {Lawson}, \citenamefont {Leder}, \citenamefont {Miyazaki}, \citenamefont {Morampudi}, \citenamefont {Peiris}, \citenamefont {R\o{}ising}, \citenamefont {Singh}, \citenamefont {Sun}, \citenamefont {Thomas}, \citenamefont {Wilczek}, \citenamefont {Withington}, \citenamefont {Wooten}, \citenamefont {Dilling}, \citenamefont {Febbraro}, \citenamefont {Knirck},\ and\ \citenamefont {Marvinney}}]{millar_2023_alpha}%
  \BibitemOpen
  \bibfield  {author} {\bibinfo {author} {\bibfnamefont {A.~J.}\ \bibnamefont {Millar}}, \bibinfo {author} {\bibfnamefont {S.~M.}\ \bibnamefont {Anlage}}, \bibinfo {author} {\bibfnamefont {R.}~\bibnamefont {Balafendiev}}, \bibinfo {author} {\bibfnamefont {P.}~\bibnamefont {Belov}}, \bibinfo {author} {\bibfnamefont {K.}~\bibnamefont {van Bibber}}, \bibinfo {author} {\bibfnamefont {J.}~\bibnamefont {Conrad}}, \bibinfo {author} {\bibfnamefont {M.}~\bibnamefont {Demarteau}}, \bibinfo {author} {\bibfnamefont {A.}~\bibnamefont {Droster}}, \bibinfo {author} {\bibfnamefont {K.}~\bibnamefont {Dunne}}, \bibinfo {author} {\bibfnamefont {A.~G.}\ \bibnamefont {Rosso}}, \bibinfo {author} {\bibfnamefont {J.~E.}\ \bibnamefont {Gudmundsson}}, \bibinfo {author} {\bibfnamefont {H.}~\bibnamefont {Jackson}}, \bibinfo {author} {\bibfnamefont {G.}~\bibnamefont {Kaur}}, \bibinfo {author} {\bibfnamefont {T.}~\bibnamefont {Klaesson}}, \bibinfo {author} {\bibfnamefont {N.}~\bibnamefont {Kowitt}}, \bibinfo {author} {\bibfnamefont
  {M.}~\bibnamefont {Lawson}}, \bibinfo {author} {\bibfnamefont {A.}~\bibnamefont {Leder}}, \bibinfo {author} {\bibfnamefont {A.}~\bibnamefont {Miyazaki}}, \bibinfo {author} {\bibfnamefont {S.}~\bibnamefont {Morampudi}}, \bibinfo {author} {\bibfnamefont {H.~V.}\ \bibnamefont {Peiris}}, \bibinfo {author} {\bibfnamefont {H.~S.}\ \bibnamefont {R\o{}ising}}, \bibinfo {author} {\bibfnamefont {G.}~\bibnamefont {Singh}}, \bibinfo {author} {\bibfnamefont {D.}~\bibnamefont {Sun}}, \bibinfo {author} {\bibfnamefont {J.~H.}\ \bibnamefont {Thomas}}, \bibinfo {author} {\bibfnamefont {F.}~\bibnamefont {Wilczek}}, \bibinfo {author} {\bibfnamefont {S.}~\bibnamefont {Withington}}, \bibinfo {author} {\bibfnamefont {M.}~\bibnamefont {Wooten}}, \bibinfo {author} {\bibfnamefont {J.}~\bibnamefont {Dilling}}, \bibinfo {author} {\bibfnamefont {M.}~\bibnamefont {Febbraro}}, \bibinfo {author} {\bibfnamefont {S.}~\bibnamefont {Knirck}},\ and\ \bibinfo {author} {\bibfnamefont {C.}~\bibnamefont {Marvinney}} (\bibinfo {collaboration}
  {Endorsers}),\ }\href {https://doi.org/10.1103/PhysRevD.107.055013} {\bibfield  {journal} {\bibinfo  {journal} {Phys. Rev. D}\ }\textbf {\bibinfo {volume} {107}},\ \bibinfo {pages} {055013} (\bibinfo {year} {2023})}\BibitemShut {NoStop}%
\bibitem [{\citenamefont {McAllister}\ \emph {et~al.}(2017)\citenamefont {McAllister}, \citenamefont {Flower}, \citenamefont {Kruger}, \citenamefont {Ivanov}, \citenamefont {Goryachev}, \citenamefont {Bourhill},\ and\ \citenamefont {Tobar}}]{McAllister_2017_organ}%
  \BibitemOpen
  \bibfield  {author} {\bibinfo {author} {\bibfnamefont {B.~T.}\ \bibnamefont {McAllister}}, \bibinfo {author} {\bibfnamefont {G.}~\bibnamefont {Flower}}, \bibinfo {author} {\bibfnamefont {J.}~\bibnamefont {Kruger}}, \bibinfo {author} {\bibfnamefont {E.~N.}\ \bibnamefont {Ivanov}}, \bibinfo {author} {\bibfnamefont {M.}~\bibnamefont {Goryachev}}, \bibinfo {author} {\bibfnamefont {J.}~\bibnamefont {Bourhill}},\ and\ \bibinfo {author} {\bibfnamefont {M.~E.}\ \bibnamefont {Tobar}},\ }\href {https://doi.org/10.1016/j.dark.2017.09.010} {\bibfield  {journal} {\bibinfo  {journal} {Phys. Dark Univ.}\ }\textbf {\bibinfo {volume} {18}},\ \bibinfo {pages} {67} (\bibinfo {year} {2017})},\ \Eprint {https://arxiv.org/abs/1706.00209} {arXiv:1706.00209 [physics.ins-det]} \BibitemShut {NoStop}%
\bibitem [{\citenamefont {Withers}\ and\ \citenamefont {Kuo}(2025)}]{Withers_2024_beehive}%
  \BibitemOpen
  \bibfield  {author} {\bibinfo {author} {\bibfnamefont {M.~O.}\ \bibnamefont {Withers}}\ and\ \bibinfo {author} {\bibfnamefont {C.-L.}\ \bibnamefont {Kuo}},\ }\href {https://doi.org/10.1103/PhysRevD.111.072011} {\bibfield  {journal} {\bibinfo  {journal} {Phys. Rev. D}\ }\textbf {\bibinfo {volume} {111}},\ \bibinfo {pages} {072011} (\bibinfo {year} {2025})},\ \Eprint {https://arxiv.org/abs/2404.06627} {arXiv:2404.06627 [hep-ex]} \BibitemShut {NoStop}%
\bibitem [{\citenamefont {Zhitnitsky}(2023)}]{Zhitnitsky_2023_structure}%
  \BibitemOpen
  \bibfield  {author} {\bibinfo {author} {\bibfnamefont {A.}~\bibnamefont {Zhitnitsky}},\ }\href {https://doi.org/https://doi.org/10.1016/j.dark.2023.101217} {\bibfield  {journal} {\bibinfo  {journal} {Physics of the Dark Universe}\ }\textbf {\bibinfo {volume} {40}},\ \bibinfo {pages} {101217} (\bibinfo {year} {2023})}\BibitemShut {NoStop}%
\bibitem [{\citenamefont {Fischer}\ \emph {et~al.}(2018)\citenamefont {Fischer}, \citenamefont {Liang}, \citenamefont {Zhitnitsky}, \citenamefont {Semertzidis},\ and\ \citenamefont {Zioutas}}]{Fischer_CAST_2018}%
  \BibitemOpen
  \bibfield  {author} {\bibinfo {author} {\bibfnamefont {H.}~\bibnamefont {Fischer}}, \bibinfo {author} {\bibfnamefont {X.}~\bibnamefont {Liang}}, \bibinfo {author} {\bibfnamefont {A.}~\bibnamefont {Zhitnitsky}}, \bibinfo {author} {\bibfnamefont {Y.}~\bibnamefont {Semertzidis}},\ and\ \bibinfo {author} {\bibfnamefont {K.}~\bibnamefont {Zioutas}},\ }\href {https://doi.org/10.1103/PhysRevD.98.043013} {\bibfield  {journal} {\bibinfo  {journal} {Phys. Rev. D}\ }\textbf {\bibinfo {volume} {98}},\ \bibinfo {pages} {043013} (\bibinfo {year} {2018})}\BibitemShut {NoStop}%
\bibitem [{\citenamefont {Sekatchev}\ \emph {et~al.}(2025)\citenamefont {Sekatchev}, \citenamefont {Liang}, \citenamefont {Majidi}, \citenamefont {Scully}, \citenamefont {Van~Waerbeke},\ and\ \citenamefont {Zhitnitsky}}]{Sekatchev_2025_glow}%
  \BibitemOpen
  \bibfield  {author} {\bibinfo {author} {\bibfnamefont {M.}~\bibnamefont {Sekatchev}}, \bibinfo {author} {\bibfnamefont {X.}~\bibnamefont {Liang}}, \bibinfo {author} {\bibfnamefont {F.}~\bibnamefont {Majidi}}, \bibinfo {author} {\bibfnamefont {B.}~\bibnamefont {Scully}}, \bibinfo {author} {\bibfnamefont {L.}~\bibnamefont {Van~Waerbeke}},\ and\ \bibinfo {author} {\bibfnamefont {A.}~\bibnamefont {Zhitnitsky}},\ }\href@noop {} {\bibinfo {title} {{The Glow of Axion Quark Nugget Dark Matter: (III) The Mysteries of the Milky Way UV Background}}} (\bibinfo {year} {2025}),\ \Eprint {https://arxiv.org/abs/2504.15382} {arXiv:2504.15382 [astro-ph.CO]} \BibitemShut {NoStop}%
\bibitem [{\citenamefont {Lawson}\ \emph {et~al.}(2019)\citenamefont {Lawson}, \citenamefont {Liang}, \citenamefont {Mead}, \citenamefont {Siddiqui}, \citenamefont {Van~Waerbeke},\ and\ \citenamefont {Zhitnitsky}}]{Lawson_2019_trapped}%
  \BibitemOpen
  \bibfield  {author} {\bibinfo {author} {\bibfnamefont {K.}~\bibnamefont {Lawson}}, \bibinfo {author} {\bibfnamefont {X.}~\bibnamefont {Liang}}, \bibinfo {author} {\bibfnamefont {A.}~\bibnamefont {Mead}}, \bibinfo {author} {\bibfnamefont {M.~S.~R.}\ \bibnamefont {Siddiqui}}, \bibinfo {author} {\bibfnamefont {L.}~\bibnamefont {Van~Waerbeke}},\ and\ \bibinfo {author} {\bibfnamefont {A.}~\bibnamefont {Zhitnitsky}},\ }\href {https://doi.org/10.1103/PhysRevD.100.043531} {\bibfield  {journal} {\bibinfo  {journal} {Phys. Rev. D}\ }\textbf {\bibinfo {volume} {100}},\ \bibinfo {pages} {043531} (\bibinfo {year} {2019})}\BibitemShut {NoStop}%
\bibitem [{\citenamefont {Zhitnitsky}\ and\ \citenamefont {Maroudas}(2025)}]{Zhitnitsky_mysterious_2025}%
  \BibitemOpen
  \bibfield  {author} {\bibinfo {author} {\bibfnamefont {A.}~\bibnamefont {Zhitnitsky}}\ and\ \bibinfo {author} {\bibfnamefont {M.}~\bibnamefont {Maroudas}},\ }\bibfield  {journal} {\bibinfo  {journal} {Symmetry}\ }\textbf {\bibinfo {volume} {17}},\ \href {https://doi.org/10.3390/sym17010079} {10.3390/sym17010079} (\bibinfo {year} {2025})\BibitemShut {NoStop}%
\bibitem [{\citenamefont {Zioutas}\ \emph {et~al.}(2020)\citenamefont {Zioutas}, \citenamefont {Argiriou}, \citenamefont {Fischer}, \citenamefont {Hofmann}, \citenamefont {Maroudas}, \citenamefont {Pappa},\ and\ \citenamefont {Semertzidis}}]{Zioutas_2020_stratosphere}%
  \BibitemOpen
  \bibfield  {author} {\bibinfo {author} {\bibfnamefont {K.}~\bibnamefont {Zioutas}}, \bibinfo {author} {\bibfnamefont {A.}~\bibnamefont {Argiriou}}, \bibinfo {author} {\bibfnamefont {H.}~\bibnamefont {Fischer}}, \bibinfo {author} {\bibfnamefont {S.}~\bibnamefont {Hofmann}}, \bibinfo {author} {\bibfnamefont {M.}~\bibnamefont {Maroudas}}, \bibinfo {author} {\bibfnamefont {A.}~\bibnamefont {Pappa}},\ and\ \bibinfo {author} {\bibfnamefont {Y.}~\bibnamefont {Semertzidis}},\ }\href {https://doi.org/https://doi.org/10.1016/j.dark.2020.100497} {\bibfield  {journal} {\bibinfo  {journal} {Physics of the Dark Universe}\ }\textbf {\bibinfo {volume} {28}},\ \bibinfo {pages} {100497} (\bibinfo {year} {2020})}\BibitemShut {NoStop}%
\bibitem [{\citenamefont {Bertolucci}\ \emph {et~al.}(2017)\citenamefont {Bertolucci}, \citenamefont {Zioutas}, \citenamefont {Hofmann},\ and\ \citenamefont {Maroudas}}]{Bertolucci_2017_sun}%
  \BibitemOpen
  \bibfield  {author} {\bibinfo {author} {\bibfnamefont {S.}~\bibnamefont {Bertolucci}}, \bibinfo {author} {\bibfnamefont {K.}~\bibnamefont {Zioutas}}, \bibinfo {author} {\bibfnamefont {S.}~\bibnamefont {Hofmann}},\ and\ \bibinfo {author} {\bibfnamefont {M.}~\bibnamefont {Maroudas}},\ }\href {https://doi.org/https://doi.org/10.1016/j.dark.2017.06.001} {\bibfield  {journal} {\bibinfo  {journal} {Physics of the Dark Universe}\ }\textbf {\bibinfo {volume} {17}},\ \bibinfo {pages} {13} (\bibinfo {year} {2017})}\BibitemShut {NoStop}%
\bibitem [{\citenamefont {Zioutas}\ \emph {et~al.}(2024)\citenamefont {Zioutas} \emph {et~al.}}]{Zioutas_2024_LHC}%
  \BibitemOpen
  \bibfield  {author} {\bibinfo {author} {\bibfnamefont {K.}~\bibnamefont {Zioutas}} \emph {et~al.},\ }\href@noop {} {\bibinfo {title} {{Search for anti-quark nuggets via their interaction with the LHC beam}}} (\bibinfo {year} {2024}),\ \Eprint {https://arxiv.org/abs/2403.05608} {arXiv:2403.05608 [hep-ex]} \BibitemShut {NoStop}%
\bibitem [{\citenamefont {Liang}\ and\ \citenamefont {Zhitnitsky}(2026)}]{Liang_UFO_2026}%
  \BibitemOpen
  \bibfield  {author} {\bibinfo {author} {\bibfnamefont {X.}~\bibnamefont {Liang}}\ and\ \bibinfo {author} {\bibfnamefont {A.}~\bibnamefont {Zhitnitsky}},\ }\href@noop {} {\bibinfo {title} {{Unidentified falling objects in the LHC as dark matter signals}}} (\bibinfo {year} {2026}),\ \Eprint {https://arxiv.org/abs/2602.10562} {arXiv:2602.10562 [hep-ph]} \BibitemShut {NoStop}%
\bibitem [{\citenamefont {{Kolb}}\ and\ \citenamefont {{Tkachev}}(1993)}]{Kolb_1993_miniclusters}%
  \BibitemOpen
  \bibfield  {author} {\bibinfo {author} {\bibfnamefont {E.~W.}\ \bibnamefont {{Kolb}}}\ and\ \bibinfo {author} {\bibfnamefont {I.~I.}\ \bibnamefont {{Tkachev}}},\ }\href {https://doi.org/10.1103/PhysRevLett.71.3051} {\bibfield  {journal} {\bibinfo  {journal} {Physical Review Letters}\ }\textbf {\bibinfo {volume} {71}},\ \bibinfo {pages} {3051} (\bibinfo {year} {1993})},\ \Eprint {https://arxiv.org/abs/hep-ph/9303313} {arXiv:hep-ph/9303313 [hep-ph]} \BibitemShut {NoStop}%
\bibitem [{\citenamefont {O'Hare}\ and\ \citenamefont {Pierobon}(2025)}]{OHare_2025_finegrained}%
  \BibitemOpen
  \bibfield  {author} {\bibinfo {author} {\bibfnamefont {C.~A.~J.}\ \bibnamefont {O'Hare}}\ and\ \bibinfo {author} {\bibfnamefont {G.}~\bibnamefont {Pierobon}},\ }\href@noop {} {\bibinfo {title} {{Fine-grained dark matter substructure and axion haloscopes}}} (\bibinfo {year} {2025}),\ \Eprint {https://arxiv.org/abs/2509.14874} {arXiv:2509.14874 [hep-ph]} \BibitemShut {NoStop}%
\bibitem [{\citenamefont {Knirck}\ \emph {et~al.}(2018)\citenamefont {Knirck}, \citenamefont {Millar}, \citenamefont {O'Hare}, \citenamefont {Redondo},\ and\ \citenamefont {Steffen}}]{Knirck_2018_directional}%
  \BibitemOpen
  \bibfield  {author} {\bibinfo {author} {\bibfnamefont {S.}~\bibnamefont {Knirck}}, \bibinfo {author} {\bibfnamefont {A.~J.}\ \bibnamefont {Millar}}, \bibinfo {author} {\bibfnamefont {C.~A.~J.}\ \bibnamefont {O'Hare}}, \bibinfo {author} {\bibfnamefont {J.}~\bibnamefont {Redondo}},\ and\ \bibinfo {author} {\bibfnamefont {F.~D.}\ \bibnamefont {Steffen}},\ }\href {https://doi.org/10.1088/1475-7516/2018/11/051} {\bibfield  {journal} {\bibinfo  {journal} {JCAP}\ }\textbf {\bibinfo {volume} {11}},\ \bibinfo {pages} {051}},\ \Eprint {https://arxiv.org/abs/1806.05927} {arXiv:1806.05927 [astro-ph.CO]} \BibitemShut {NoStop}%
\bibitem [{\citenamefont {Zioutas}\ \emph {et~al.}(2017)\citenamefont {Zioutas} \emph {et~al.}}]{Zioutas_2017_search}%
  \BibitemOpen
  \bibfield  {author} {\bibinfo {author} {\bibfnamefont {K.}~\bibnamefont {Zioutas}} \emph {et~al.},\ }\href@noop {} {\bibinfo {title} {{Search for axions in streaming dark matter}}} (\bibinfo {year} {2017}),\ \Eprint {https://arxiv.org/abs/1703.01436} {arXiv:1703.01436 [physics.ins-det]} \BibitemShut {NoStop}%
\bibitem [{\citenamefont {Sofue}(2020)}]{Sofue_2020_focusing}%
  \BibitemOpen
  \bibfield  {author} {\bibinfo {author} {\bibfnamefont {Y.}~\bibnamefont {Sofue}},\ }\href {https://doi.org/10.3390/galaxies8020042} {\bibfield  {journal} {\bibinfo  {journal} {Galaxies}\ }\textbf {\bibinfo {volume} {8}},\ \bibinfo {pages} {42} (\bibinfo {year} {2020})},\ \Eprint {https://arxiv.org/abs/2005.08252} {arXiv:2005.08252 [astro-ph.HE]} \BibitemShut {NoStop}%
\bibitem [{\citenamefont {Di~Vora}\ \emph {et~al.}(2025)\citenamefont {Di~Vora}, \citenamefont {Lombardi}, \citenamefont {Ortolan}, \citenamefont {Ruoso}, \citenamefont {Braggio}, \citenamefont {Carugno},\ and\ \citenamefont {Gardikiotis}}]{DiVora_2025_axion}%
  \BibitemOpen
  \bibfield  {author} {\bibinfo {author} {\bibfnamefont {R.}~\bibnamefont {Di~Vora}}, \bibinfo {author} {\bibfnamefont {A.}~\bibnamefont {Lombardi}}, \bibinfo {author} {\bibfnamefont {A.}~\bibnamefont {Ortolan}}, \bibinfo {author} {\bibfnamefont {G.}~\bibnamefont {Ruoso}}, \bibinfo {author} {\bibfnamefont {C.}~\bibnamefont {Braggio}}, \bibinfo {author} {\bibfnamefont {G.}~\bibnamefont {Carugno}},\ and\ \bibinfo {author} {\bibfnamefont {A.}~\bibnamefont {Gardikiotis}},\ }\href {https://doi.org/10.1103/PhysRevApplied.23.034047} {\bibfield  {journal} {\bibinfo  {journal} {Phys. Rev. Appl.}\ }\textbf {\bibinfo {volume} {23}},\ \bibinfo {pages} {034047} (\bibinfo {year} {2025})}\BibitemShut {NoStop}%
\bibitem [{\citenamefont {Cezar}\ \emph {et~al.}(2026)\citenamefont {Cezar}, \citenamefont {Maroudas},\ and\ \citenamefont {Horns}}]{Cezar_2026_modular}%
  \BibitemOpen
  \bibfield  {author} {\bibinfo {author} {\bibfnamefont {T.-S.}\ \bibnamefont {Cezar}}, \bibinfo {author} {\bibfnamefont {M.}~\bibnamefont {Maroudas}},\ and\ \bibinfo {author} {\bibfnamefont {D.}~\bibnamefont {Horns}},\ }\href@noop {} {\bibinfo {title} {{A Modular Zero-Dead-Time Data Acquisition and Real-Time GPU Processing Platform for High Throughput Physics Experiments}}} (\bibinfo {year} {2026}),\ \Eprint {https://arxiv.org/abs/2605.08845} {arXiv:2605.08845 [physics.ins-det]} \BibitemShut {NoStop}%
\bibitem [{\citenamefont {D{\'\i}az-Morcillo}\ \emph {et~al.}(2022)\citenamefont {D{\'\i}az-Morcillo}, \citenamefont {Garc{\'\i}a~Barcel{\'o}}, \citenamefont {Lozano~Guerrero}, \citenamefont {Navarro}, \citenamefont {Gimeno}, \citenamefont {Arguedas~Cuendis}, \citenamefont {{\'A}lvarez~Melc{\'o}n}, \citenamefont {Cogollos}, \citenamefont {Calatroni}, \citenamefont {D{\"o}brich}, \citenamefont {Gallego-Puyol}, \citenamefont {Golm}, \citenamefont {Irastorza}, \citenamefont {Malbrunot}, \citenamefont {Miralda-Escud{\'e}}, \citenamefont {Pe{\~n}a~Garay}, \citenamefont {Redondo},\ and\ \citenamefont {Wuensch}}]{rades_2022_design}%
  \BibitemOpen
  \bibfield  {author} {\bibinfo {author} {\bibfnamefont {A.}~\bibnamefont {D{\'\i}az-Morcillo}}, \bibinfo {author} {\bibfnamefont {J.~M.}\ \bibnamefont {Garc{\'\i}a~Barcel{\'o}}}, \bibinfo {author} {\bibfnamefont {A.~J.}\ \bibnamefont {Lozano~Guerrero}}, \bibinfo {author} {\bibfnamefont {P.}~\bibnamefont {Navarro}}, \bibinfo {author} {\bibfnamefont {B.}~\bibnamefont {Gimeno}}, \bibinfo {author} {\bibfnamefont {S.}~\bibnamefont {Arguedas~Cuendis}}, \bibinfo {author} {\bibfnamefont {A.}~\bibnamefont {{\'A}lvarez~Melc{\'o}n}}, \bibinfo {author} {\bibfnamefont {C.}~\bibnamefont {Cogollos}}, \bibinfo {author} {\bibfnamefont {S.}~\bibnamefont {Calatroni}}, \bibinfo {author} {\bibfnamefont {B.}~\bibnamefont {D{\"o}brich}}, \bibinfo {author} {\bibfnamefont {J.~D.}\ \bibnamefont {Gallego-Puyol}}, \bibinfo {author} {\bibfnamefont {J.}~\bibnamefont {Golm}}, \bibinfo {author} {\bibfnamefont {I.~G.}\ \bibnamefont {Irastorza}}, \bibinfo {author} {\bibfnamefont {C.}~\bibnamefont {Malbrunot}}, \bibinfo {author}
  {\bibfnamefont {J.}~\bibnamefont {Miralda-Escud{\'e}}}, \bibinfo {author} {\bibfnamefont {C.}~\bibnamefont {Pe{\~n}a~Garay}}, \bibinfo {author} {\bibfnamefont {J.}~\bibnamefont {Redondo}},\ and\ \bibinfo {author} {\bibfnamefont {W.}~\bibnamefont {Wuensch}},\ }\bibfield  {journal} {\bibinfo  {journal} {Universe}\ }\textbf {\bibinfo {volume} {8}},\ \href {https://doi.org/10.3390/universe8010005} {10.3390/universe8010005} (\bibinfo {year} {2022})\BibitemShut {NoStop}%
\bibitem [{\citenamefont {O'Hare}(2020)}]{OHare_2020_axionlimits}%
  \BibitemOpen
  \bibfield  {author} {\bibinfo {author} {\bibfnamefont {C.}~\bibnamefont {O'Hare}},\ }\href {https://doi.org/10.5281/zenodo.3932430} {\bibinfo {title} {{cajohare/axionlimits: Axionlimits (v1.0)}}} (\bibinfo {year} {2020})\BibitemShut {NoStop}%
\bibitem [{\citenamefont {Ling}\ \emph {et~al.}(2004)\citenamefont {Ling}, \citenamefont {Sikivie},\ and\ \citenamefont {Wick}}]{Ling_2004_diurnal}%
  \BibitemOpen
  \bibfield  {author} {\bibinfo {author} {\bibfnamefont {F.-S.}\ \bibnamefont {Ling}}, \bibinfo {author} {\bibfnamefont {P.}~\bibnamefont {Sikivie}},\ and\ \bibinfo {author} {\bibfnamefont {S.}~\bibnamefont {Wick}},\ }\href {https://doi.org/10.1103/PhysRevD.70.123503} {\bibfield  {journal} {\bibinfo  {journal} {Phys. Rev. D}\ }\textbf {\bibinfo {volume} {70}},\ \bibinfo {pages} {123503} (\bibinfo {year} {2004})}\BibitemShut {NoStop}%
\bibitem [{\citenamefont {Gramolin}\ \emph {et~al.}(2022)\citenamefont {Gramolin}, \citenamefont {Wickenbrock}, \citenamefont {Aybas}, \citenamefont {Bekker}, \citenamefont {Budker}, \citenamefont {Centers}, \citenamefont {Figueroa}, \citenamefont {Jackson~Kimball},\ and\ \citenamefont {Sushkov}}]{Gramolin_2022_spectral}%
  \BibitemOpen
  \bibfield  {author} {\bibinfo {author} {\bibfnamefont {A.~V.}\ \bibnamefont {Gramolin}}, \bibinfo {author} {\bibfnamefont {A.}~\bibnamefont {Wickenbrock}}, \bibinfo {author} {\bibfnamefont {D.}~\bibnamefont {Aybas}}, \bibinfo {author} {\bibfnamefont {H.}~\bibnamefont {Bekker}}, \bibinfo {author} {\bibfnamefont {D.}~\bibnamefont {Budker}}, \bibinfo {author} {\bibfnamefont {G.~P.}\ \bibnamefont {Centers}}, \bibinfo {author} {\bibfnamefont {N.~L.}\ \bibnamefont {Figueroa}}, \bibinfo {author} {\bibfnamefont {D.~F.}\ \bibnamefont {Jackson~Kimball}},\ and\ \bibinfo {author} {\bibfnamefont {A.~O.}\ \bibnamefont {Sushkov}},\ }\href {https://doi.org/10.1103/PhysRevD.105.035029} {\bibfield  {journal} {\bibinfo  {journal} {Phys. Rev. D}\ }\textbf {\bibinfo {volume} {105}},\ \bibinfo {pages} {035029} (\bibinfo {year} {2022})}\BibitemShut {NoStop}%
\bibitem [{\citenamefont {{Kim}}(1979)}]{KSVZ1}%
  \BibitemOpen
  \bibfield  {author} {\bibinfo {author} {\bibfnamefont {J.~E.}\ \bibnamefont {{Kim}}},\ }\href {https://doi.org/10.1103/PhysRevLett.43.103} {\bibfield  {journal} {\bibinfo  {journal} {Physical Review Letters}\ }\textbf {\bibinfo {volume} {43}},\ \bibinfo {pages} {103} (\bibinfo {year} {1979})}\BibitemShut {NoStop}%
\bibitem [{\citenamefont {{Shifman}}\ \emph {et~al.}(1980)\citenamefont {{Shifman}}, \citenamefont {{Vainshtein}},\ and\ \citenamefont {{Zakharov}}}]{KSVZ2}%
  \BibitemOpen
  \bibfield  {author} {\bibinfo {author} {\bibfnamefont {M.~A.}\ \bibnamefont {{Shifman}}}, \bibinfo {author} {\bibfnamefont {A.~I.}\ \bibnamefont {{Vainshtein}}},\ and\ \bibinfo {author} {\bibfnamefont {V.~I.}\ \bibnamefont {{Zakharov}}},\ }\href {https://doi.org/10.1016/0550-3213(80)90209-6} {\bibfield  {journal} {\bibinfo  {journal} {Nucl. Phys. B}\ }\textbf {\bibinfo {volume} {166}},\ \bibinfo {pages} {493} (\bibinfo {year} {1980})}\BibitemShut {NoStop}%
\bibitem [{\citenamefont {{Dine}}\ \emph {et~al.}(1981)\citenamefont {{Dine}}, \citenamefont {{Fischler}},\ and\ \citenamefont {{Srednicki}}}]{DFSZ1}%
  \BibitemOpen
  \bibfield  {author} {\bibinfo {author} {\bibfnamefont {M.}~\bibnamefont {{Dine}}}, \bibinfo {author} {\bibfnamefont {W.}~\bibnamefont {{Fischler}}},\ and\ \bibinfo {author} {\bibfnamefont {M.}~\bibnamefont {{Srednicki}}},\ }\href {https://doi.org/10.1016/0370-2693(81)90590-6} {\bibfield  {journal} {\bibinfo  {journal} {Phys. Lett. B}\ }\textbf {\bibinfo {volume} {104}},\ \bibinfo {pages} {199} (\bibinfo {year} {1981})}\BibitemShut {NoStop}%
\bibitem [{\citenamefont {Zhitnitsky}(1980)}]{DFSZ2}%
  \BibitemOpen
  \bibfield  {author} {\bibinfo {author} {\bibfnamefont {A.~R.}\ \bibnamefont {Zhitnitsky}},\ }\href@noop {} {\bibfield  {journal} {\bibinfo  {journal} {Sov. J. Nucl. Phys.}\ }\textbf {\bibinfo {volume} {31}},\ \bibinfo {pages} {260} (\bibinfo {year} {1980})},\ \bibinfo {note} {[Yad. Fiz.31,497(1980)]}\BibitemShut {NoStop}%
\end{thebibliography}%

\end{document}